\documentclass[aps,prd,superscriptaddress,amsmath,amsfont,amssymb,preprint,nofootinbib]{revtex4}
\pdfoutput=1
\allowdisplaybreaks
\usepackage{bbold}
\usepackage{subfigure}
\usepackage{slashed}
\usepackage{mathtools}
\usepackage[english]{babel}
\usepackage{mathrsfs}

\newcommand{\C}{{\cal C}}
\newcommand{\Q}{{\cal Q}}

\newcommand{\lag}{{\mathscr{L}}}
\newcommand{\todo}[1]{{\color{red} \ifmmode\else[todo]\fi #1}}
\usepackage[usenames,dvipsnames]{xcolor}
     \definecolor{hgreen}{rgb}{0,.3,0}
     \definecolor{hred}{rgb}{.3,0,0}
     \definecolor{hblue}{rgb}{0,0,.3}
     \definecolor{LightGray}{gray}{0.95}
\usepackage[backref=page,
            colorlinks=true,
            linkcolor=hblue,
            citecolor=hgreen,
            filecolor=hblue,
            urlcolor=hred]{hyperref}

\usepackage{fancyvrb}
\usepackage{upquote}

\renewcommand*{\backref}[1]{}

\newcommand{\beq}{\begin{equation}}
\newcommand{\eeq}{\end{equation}} 
\newcommand{\bi}{\begin{itemize}}
\newcommand{\ei}{\end{itemize}}

\definecolor{Red}{rgb}{1.,0.,0.}
\definecolor{Grn}{rgb}{0.,0.75,0.}
\definecolor{Blu}{rgb}{0.,0.,1.}
\definecolor{DrkGrn}{HTML}{00AA00}
\definecolor{Maroon}{HTML}{800000}

\DeclareMathOperator{\diag}{diag}

\definecolor{pan624}{rgb}{0.482,0.635,0.588} 
\definecolor{pan576}{rgb}{0.412,0.569,0.231} 
\definecolor{pan129}{rgb}{0.961,0.812,0.278}
\definecolor{pan5405}{rgb}{0,0.129,0.278} 
\definecolor{shadecolor}{rgb}{0.482,0.635,0.588}
\definecolor{mygray}{HTML}{666666}
\definecolor{x11steelblue}{HTML}{4682B4}
\definecolor{x11firebrick}{HTML}{B22222}
\definecolor{x11forestgreen}{HTML}{228B22}

\newcommand{\muew}{\mu_{\rm ew}}
\newcommand{\mustr}{\mu_{\rm had}}

\begin{document}

\title{Weak mixing below the weak scale in dark-matter direct detection}

\def\Cincy{Department of Physics, University of Cincinnati, Cincinnati, Ohio 45221,USA}
\def\UCSD{Department of Physics, University of California-San Diego, La Jolla, CA 92093, USA}
\def\TUD{Fakult\"at f\"ur Physik, TU Dortmund, D-44221 Dortmund, Germany} 
\def\Chicago{Enrico Fermi Institute, University of Chicago, Chicago, IL 60637, USA}

\author{\textbf{Joachim Brod}}
\email{joachim.brod@tu-dortmund.de}
\affiliation{\TUD}

 \author{\textbf{Benjamin Grinstein}}
 \email{bgrinstein@ucsd.edu}
 \affiliation{\UCSD}

\author{\textbf{Emmanuel Stamou}}
\email{estamou@uchicago.edu}
\affiliation{\Chicago}

\author{\textbf{Jure Zupan}} 
\email{zupanje@ucmail.uc.edu}
\affiliation{\Cincy}

\date{\today}

\begin{abstract}
If dark matter couples predominantly to the axial-vector currents with
heavy quarks, the leading contribution to dark-matter scattering on
nuclei is either due to one-loop weak corrections or due to the
heavy-quark axial charges of the nucleons.  We calculate the effects
of Higgs and weak gauge-boson exchanges for dark matter coupling to
heavy-quark axial-vector currents in an effective theory below the
weak scale. By explicit computation, we show that the
leading-logarithmic QCD corrections are important, and thus resum them
to all orders using the renormalization group.
\end{abstract}

\preprint{DO-TH 17/20}

\maketitle
\tableofcontents

\section{Introduction\label{sec:Intro}}

A useful approach to describe the results of Dark Matter (DM)
direct-detection experiments is to relate them to an Effective Field
Theory (EFT) of DM coupling to quarks, gluons, leptons, and photons
\cite{Bagnasco:1993st, Kurylov:2003ra, Kopp:2009qt, Goodman:2010qn,
  Goodman:2010ku, Bai:2010hh, Hill:2011be, DelNobile:2013sia,
  Hill:2014yxa, Crivellin:2014qxa, DEramo:2014nmf, Hill:2013hoa,
  Hoferichter:2015ipa, Bishara:2016hek, DEramo:2016gos,
  Bishara:2017pfq, DEramo:2017zqw}.  In this EFT, the level of
suppression of DM interactions with the Standard Model (SM) depends on
the mass dimension of the interaction operators, i.e., the higher the
mass dimension the more suppressed the operator is. The mass dimension
of operators is thus the organizing principle in capturing the
phenomenologically most relevant effects, which is why in
phenomenological analyses one keeps all relevant terms up to some mass
dimension, $d$.  An important question is, at which value of $d$ one
can truncate the expansion.  The obvious choice would be to keep all
operators of dimension five and six, and a subset of dimension-seven
operators that do not involve derivatives, as in this case one covers
most of the UV models of DM.

In this work, we show that the leading contribution to the scattering
cross section originates from double insertions of dimension-six
operators if the DM interaction is predominantly due to DM vector
currents coupling to heavy-quark axial-vector currents. This
effectively means that in such a case it is necessary to extend the
EFT to include operators of mass dimension eight. That such
corrections are important was first pointed out in
Refs.~\cite{Crivellin:2014qxa, DEramo:2014nmf}, with the
phenomenological implications further discussed in
\cite{DEramo:2017zqw}.  We improve on the analysis of
Ref.~\cite{DEramo:2014nmf} in two ways: {\it i)} we clarify how to
consistently include the double-insertion contributions in the EFT
framework, {\it ii)} we also perform the resummation of the QCD
corrections at leading-logarithmic accuracy. Moreover, the generality
of our approach covers also the case of non-singlet DM in the theory
above the electroweak scale.

The paper is structured as follows. In Sections
\ref{sec:power:counting}--\ref{sec:UV} we derive our results for the
case of Dirac-fermion DM. These are then extended to the case of
Majorana-fermion DM and to the case of scalar DM in Section
\ref{sec:other}.  In Section \ref{sec:power:counting} we first show
that the electroweak corrections have to be included if DM couples
only to vector or axial-vector currents with heavy quarks. The weak
interactions below the weak scale are encoded in an effective
Lagrangian, which is introduced in Section \ref{sec:SMweak}.  Section
\ref{sec:RGE} contains our results for the anomalous dimensions
controlling the operator mixing, while the renormalization-group
evolution is given in Section \ref{sec:match}.  In Section
\ref{sec:UV} we show how our results connect to the physics above the
electroweak scale. Section \ref{sec:discussion} contains conclusions,
while Appendix \ref{App:unphysical} collects some unphysical operators
entering in intermediate steps of our calculation.

\section{The importance of weak corrections for axial currents\label{sec:power:counting}}

We start by considering the DM EFT valid below the electroweak scale,
$\mu_b< \mu< \muew$, for Dirac-fermion DM when five quark flavors are
active,
\begin{equation}\label{eq:d6:DM:Lf5}
\lag_\chi=\sum_{a,d} \frac{\C_{a}^{(d)}}{\Lambda^{d-4}} \Q_a^{(d)}\,.
\end{equation}
The sums run over the dimensions of the operators, $d$, and the
operator labels, $a$.  The operators are multiplied by dimensionless
Wilson coefficients, $\C_{a}^{(d)}$, and the appropriate powers of the
mediator mass scale, $\Lambda$. Since we are interested in the theory
below the electroweak scale, any interactions with the top quark, $W$,
$Z$ bosons, and the Higgs are integrated out and are part of the
Wilson coefficients $\C_a^{(d)}$.  In this work, we focus on
dimension-six operators, namely
\begin{align}
{\cal Q}_{1,f}^{(6)} & = (\bar \chi \gamma_\mu \chi) (\bar f \gamma^\mu f)\,,&
{\cal Q}_{2,f}^{(6)} &= (\bar \chi\gamma_\mu\gamma_5 \chi)(\bar f \gamma^\mu f)\,, \label{eq:dim6:Q1Q2}\\ 
{\cal Q}_{3,f}^{(6)} & = (\bar \chi \gamma_\mu \chi)(\bar f \gamma^\mu \gamma_5 f)\,,& 
{\cal Q}_{4,f}^{(6)} & = (\bar \chi\gamma_\mu\gamma_5 \chi)(\bar f \gamma^\mu \gamma_5 f)\,,\label{eq:dim6:Q3Q4}
\end{align}
where $f$ can be any of the SM fermions apart from the top quark.  Our
dimension counting follows
Refs.~\cite{Bishara:2017pfq,Bishara:2017nnn}, such that scalar
four-fermion operators are considered to be dimension seven, i.e., we
assume they originate from a Higgs field insertion above the
electroweak scale.

As we show below, a proper description of DM scattering on nuclei due
to dimension-six operators requires including corrections from QED and
the weak interactions. By contrast, such corrections are always
subleading for dimension-five and dimension-seven operators. The basis
of dimension-five operators, which couple DM to photons, can be found,
e.g., in Refs.~\cite{Bishara:2017pfq,Bishara:2017nnn}, while the full
basis of dimension-seven operators was derived in
Ref.~\cite{Brod:2017bsw}.

If only a single operator in
Eqs.~\eqref{eq:dim6:Q1Q2}--\eqref{eq:dim6:Q3Q4} contributes, the cross
section for DM--nucleus scattering can be written as
\begin{equation} \sigma\propto
\left(\frac{{\cal C}_{a}^{(d)} }{\Lambda^{d-4}}{\cal A}[{\cal Q}_a^{(d)}]\right)^2,
\end{equation} 
where ${\cal A}[{\cal Q}_a^{(d)}]$ is an ``effective scattering
amplitude''.  It is a product of the scattering amplitude, the nuclear
response functions \cite{Fitzpatrick:2012ix, Fitzpatrick:2012ib,
  Anand:2013yka, Hoferichter:2015ipa, Hoferichter:2016nvd,
  Bishara:2016hek}, and all the relevant kinematic factors.  We
estimate ${\cal A}[{\cal Q}_a^{(d)}]$ in three different limits: 
{\it  i)} in the limit of only strong interactions, {\it ii)} including
QED corrections, and {\it iii)} also including corrections from weak
interactions.

\bigskip

{\it i)} Switching off QED and weak interactions, the effective
scattering amplitudes for dimension-six operators have the following
parametric sizes (see Ref.~\cite{Bishara:2017nnn}):
\begin{align}
\label{eq:Q1q6:noRG}
{\cal A}[\Q_{1,u(d)}^{(6)}]&\sim A \,, & 
{\cal A}[\Q_{1,s}^{(6)}]&= 0\,,     &
{\cal A}[\Q_{1,c(b)}^{(6)}]&= 0\,,
\\
\label{eq:Q2q6:noRG}
{\cal A}[\Q_{2,u(d)}^{(6)}]&\sim {\rm max}\left\{v_T A, \frac{q}{m_N}\right\}\,, &
{\cal A}[\Q_{2,s}^{(6)}]&= 0\,, &
{\cal A}[\Q_{2,c(b)}^{(6)}]&= 0\,,  
\\
\label{eq:Q3q6:noRG}
{\cal A}[\Q_{3,u(d)}^{(6)}]&\sim {\rm max}\left\{v_T, \frac{q}{m_\chi}\right\}\,, &
{\cal A}[\Q_{3,s}^{(6)}]&\sim \Delta s  {\cal A}[{\cal Q}_{3,q}^{(6)}]\,, &
{\cal A}[\Q_{3,c(b)}^{(6)}]&\sim \Delta c (b)  {\cal A}[{\cal Q}_{3,q}^{(6)}]\,,
\\
\label{eq:Q4q6:noRG}
{\cal A}[\Q_{4,u(d)}^{(6)}]&\sim  1\,, &
{\cal A}[\Q_{4,s}^{(6)}]&\sim \Delta s  {\cal A}[{\cal Q}_{4,q}^{(6)}]\,, &
{\cal A}[\Q_{4,c(b)}^{(6)}]&\sim  \Delta c(b)  {\cal A}[{\cal Q}_{4,q}^{(6)}]\,,
\end{align}
where in the subscript $q=u,d$.  Here, $v_T\sim 10^{-3}$ is the
typical DM velocity in the laboratory frame, $q$ is the typical
momentum exchange, $q/m_N\lesssim 0.1$, where $m_N$ is the nucleon
mass, and $A$ is the nuclear mass number (for heavy nuclei $A\sim
10^2$).  The approximate expressions for the effective scattering
amplitudes in Eqs.~\eqref{eq:Q1q6:noRG}--\eqref{eq:Q4q6:noRG} include
the parametric ${\mathcal O}(A)$ coherent enhancement of the
spin-independent nuclear response function, $W_M(q)$, while all the
other response functions were counted as ${\mathcal O}(1)$. The vector
and axial form factors at zero recoil are ${\mathcal O}(1)$ for $u,d$
quarks.  For the strange, charm and bottom quarks the vector form
factors vanish.  The axial charge for the strange quark is reasonably
well known, $\Delta s=-0.031(5)$ \cite{Bishara:2017pfq, QCDSF:2011aa,
  Engelhardt:2012gd, Bhattacharya:2015gma, Alexandrou:2017hac}.  The
axial charges of charm and bottom quarks currently have a much larger
uncertainty.  Ref.~\cite{Polyakov:1998rb} obtained $\Delta c\approx -
5 \cdot 10^{-4}$, $\Delta b\approx - 5 \cdot 10^{-5}$, with probably
at least a factor of two uncertainty on these estimates.

Due to the non-relativistic nature of the problem and the sizes of the
nuclear matrix elements, there are large hierarchies between the
effective scattering amplitudes.  For light quarks this hierarchy can
be as large as $v_T/A\sim 10^{-5}$.  For heavy quarks the effective
amplitudes either vanish or are very small.  This indicates that
subleading corrections from QED and weak interactions may be
important.

\begin{figure}[t]
\includegraphics{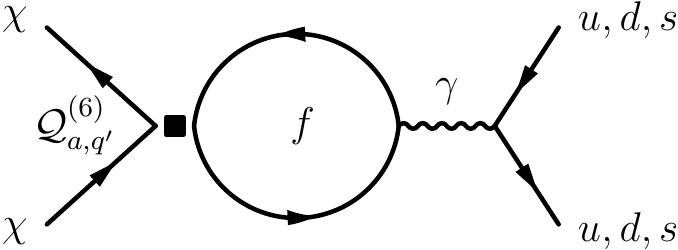}
\caption{Photon penguin diagrams that induce mixing of DM vector
  interactions with leptons or heavy quarks into DM vector
  interactions with light quarks.
  Here, $f=u,d,s,c,b,e,\mu,\tau$ can denote any of the 
  quarks or charged leptons.
\label{fig:photon-penguin}
}
\end{figure}

\bigskip

{\it ii)} We now switch on corrections due to QED interactions. The
diagrams with a closed fermion loop and a photon exchange, see
Fig.~\ref{fig:photon-penguin}, generate couplings to light quarks for
all DM interactions with quark and lepton vector currents.  The
parametric estimates in
Eqs.~\eqref{eq:Q1q6:noRG}--\eqref{eq:Q2q6:noRG} are therefore modified
to
\begin{align}
{\cal A}[\Q_{1,u(d)}^{(6)}]&\sim A\,,&
{\cal A}[\Q_{1,s}^{(6)}]&\sim \frac{\alpha}{4\pi} {\cal A}[{\cal Q}_{1,q}^{(6)}]\,,&
{\cal A}[\Q_{1,c(b)}^{(6)}]&\sim \frac{\alpha}{4\pi} {\cal A}[{\cal Q}_{1,q}^{(6)}]\,,\\
{\cal A}[\Q_{2,u(d)}^{(6)}]&\sim {\rm max}\left\{v_T A, \frac{q}{m_N}\right\}\,,&
{\cal A}[\Q_{2,s}^{(6)}]&\sim \frac{\alpha}{4\pi} {\cal A}[{\cal Q}_{2,q}^{(6)}]\,,&
{\cal A}[\Q_{2,c(b)}^{(6)}]&\sim \frac{\alpha}{4\pi} {\cal A}[{\cal Q}_{2,q}^{(6)}], 
\end{align}
while the parametric estimates for the operators with quark axial-vector 
currents, i.e., Eqs.~\eqref{eq:Q3q6:noRG}--\eqref{eq:Q4q6:noRG}, are not modified 
by the presence of QED corrections.  

\bigskip

{\it iii)} Potentially important corrections to the effective amplitudes for operators with
heavy-quark axial currents in
Eqs.~\eqref{eq:Q3q6:noRG}--\eqref{eq:Q4q6:noRG} are induced once weak
interactions are included. Below the weak scale the $W$ and $Z$ bosons
are integrated out, generating an effective weak Lagrangian
${\lag}_\text{eff}^{\rm SM}$ composed of dimension-six four-fermion
operators, see Eq.~\eqref{eq:Leff} in the next section.  A double
insertion of one four-fermion operator from ${\lag}_\text{eff}^{\rm
  SM}$ and one from $\lag_\chi$, see Fig.~\ref{fig:below_ew_fish},
induces the additional contributions to 
\begin{equation}
\label{eq:AQ34}
{\cal A}[\Q_{3,c(b)}^{(6)}]\sim \frac{\alpha}{4\pi s_w^2} \frac{m_{c(b)}^2}{m_Z^2} A\,, \qquad 
{\cal A}[\Q_{4,c(b)}^{(6)}]\sim \frac{\alpha}{4\pi s_w^2} \frac{m_{c(b)}^2}{m_Z^2} {\rm max} \left\{v_T A, \frac{q}{m_N}\right\}\,,
\end{equation}
where $s_w$ is the sine of the weak mixing angle. The proportionality
to the square of the heavy-quark mass $m_{c(b)}$ --- necessary for
dimensional reasons --- can be deduced from the fact that it is the
only relevant mass scale in the regime $\mu_{c(b)}< \mu < \muew$.  For
$\Q_{3,c(b)}^{(6)}$ these contributions dominate over the axial charge
contribution, Eq. \eqref{eq:Q3q6:noRG}, by several orders of
magnitude, while for $\Q_{4,c(b)}^{(6)}$ the electroweak corrections
are either comparable or smaller than in
Eq.~\eqref{eq:Q4q6:noRG}. More details follow in the next sections.

\bigskip

The above estimates show that QED and weak corrections are essential
to capture the leading contributions for the dimension-six operators
in Eqs.~\eqref{eq:dim6:Q1Q2}--\eqref{eq:dim6:Q3Q4} that involve heavy
quarks.  The same type of QED and weak radiative corrections also
induce the leading effective amplitudes for the scattering on nucleons
when the DM couples, at tree level, only to leptons. The
logarithmically enhanced QED contributions are known, see for instance
Refs.~\cite{Hill:2014yxa, DEramo:2014nmf, Bishara:2017nnn}. In the
present work, we calculate the logarithmically enhanced contributions
due to the weak interactions. They arise, via double insertions, at
second order in the dimension-six effective interactions,
cf.~Eq.~\eqref{eq:AQ34}. Accordingly, they can mix into
dimension-eight operators, which, therefore, also have to be included.

It turns out that the weak corrections are numerically irrelevant for
operators coupling DM to light quarks at tree level. Since the weak
interactions do not conserve parity, they can lift the velocity
suppression in the matrix elements of $Q_{3,q}^{(6)}$ through the
mixing into the coherently enhanced operator $Q_{1,q}^{(6)}$. However,
the resulting relative enhancement of order $A/v_T\sim 10^5$ is not
enough to compensate for the large suppression of the weak
corrections, of order $\alpha/(4\pi s_w^2)(m_q/m_Z)^2\lesssim 10^{-9}
(m_q/100{\rm ~MeV})^2$.

The weak corrections are also much less important for the
dimension-five and dimension-seven operators coupling DM to the SM
fields~\cite{Bishara:2017nnn, Brod:2017bsw}. Most of these operators
have a nonzero nucleon matrix element already without including
electroweak corrections, in which case the latter only give subleading
corrections. This is the case for the operators coupling DM to gluons
or photons, for pseudoscalar currents with light quarks, and for
scalar quark currents, including the ones with heavy bottom and charm
quarks.  In the special case where DM couples only to pseudoscalar
heavy-quark currents the nuclear matrix elements vanish. This remains
true also after one-loop electroweak corrections are included.

In the next two sections, we will obtain the leading-logarithmic
expressions for the electroweak contributions in Eq.~\eqref{eq:AQ34}
and also resum the QCD corrections by performing the RG running from
the weak scale, $\muew \sim {\mathcal O}(m_Z)$, to the hadronic scale,
$\mustr \sim {\mathcal O}(2{\rm ~GeV})$, where we match to the
nonrelativistic theory.

\section{Standard Model weak effective Lagrangian\label{sec:SMweak}}

The SM interactions below the weak scale are described by an effective
Lagrangian, obtained by integrating out the top quark and the $Z$,
$W$, and Higgs bosons at the scale $\muew \sim m_Z$.  In this section we
focus on quark interactions.  We discuss leptons in
Section~\ref{sec:UV}.  We can neglect any operators
involving flavor-changing neutral currents as well as terms suppressed
by off-diagonal Cabibbo--Kobayashi--Maskawa (CKM) matrix elements. The
only necessary operators are
\begin{equation} \label{eq:Leff}
\begin{split}
{\lag}_\text{eff}^{\rm SM} \supset - \sqrt{2} G_F \, \bigg \{ & \sum_{q\neq q'}
\bigg[ \frac{1}{2} \sum_{i=1,2,4,5} {\cal D}_{i,qq'}^{(6)} {\cal
    O}_{i,qq'}^{(6)} + \sum_{i=3,6} {\cal D}_{i,qq'}^{(6)} {\cal
    O}_{i,qq'}^{(6)} \bigg] + \sum_q\!\!\sum_{i=1,\ldots,4} {\cal
  D}_{i,q}^{(6)} {\cal O}_{i,q}^{(6)} \bigg \} \,,
\end{split}
\end{equation}
where $G_F$ is the Fermi constant and ${\cal D}_a^{(6)}$ are
dimensionless Wilson coefficients.  The sums run over all light
quarks, $q,q' = u,d,s,c,b$, and the labels of the operators with two
different quark flavors ($q \neq q'$)
\begin{align} 
{\cal O}_{1,qq'}^{(6)} & = (\bar q \gamma_\mu q) \, (\bar q' \gamma^\mu q') \,, 
&{\cal O}_{2,qq'}^{(6)} & = (\bar q \gamma_\mu \gamma_5 q) \, (\bar q' \gamma^\mu \gamma_5 q') \,, \label{eq:op:qq':12} \\[0.5em]
{\cal O}_{3,qq'}^{(6)} & = (\bar q \gamma_\mu \gamma_5 q) \, (\bar q' \gamma^\mu q') \,, 
&{\cal O}_{4,qq'}^{(6)} & = (\bar q \gamma_\mu \, T^a q) \, (\bar q' \gamma^\mu \, T^a q') \,, \label{eq:op:qq':34} \\[0.5em]
{\cal O}_{5,qq'}^{(6)} & = (\bar q \gamma_\mu \gamma_5 \, T^a q) \, (\bar q' \gamma^\mu \gamma_5 \, T^a q') \,, 
&{\cal O}_{6,qq'}^{(6)} & = (\bar q \gamma_\mu \gamma_5 \, T^a q) \, (\bar q' \gamma^\mu \, T^a q') \,, \label{eq:op:qq':56} 
\intertext{and a single quark flavor,}
{\cal O}_{1,q}^{(6)} & = (\bar q \gamma_\mu q) \, (\bar q \gamma^\mu q) \,, 
&{\cal O}_{2,q}^{(6)} & = (\bar q \gamma_\mu \gamma_5 q) \, (\bar q \gamma^\mu \gamma_5 q) \,, \label{eq:op:q:12}\\[0.5em]
{\cal O}_{3,q}^{(6)} & = (\bar q \gamma_\mu \gamma_5 q) \, (\bar q \gamma^\mu q) \,, 
&{\cal O}_{4,q}^{(6)} & = (\bar q \gamma_\mu \, T^a q) \, (\bar q \gamma^\mu \, T^a q) \,. \label{eq:op:q:34}
\end{align}
Here, $T^a$ are the $SU(3)_c$ generators normalised as ${\rm
  Tr}(T^aT^b)= \frac{1}{2}\delta^{ab}$.  As seen from the above
operator basis, there are fewer linearly independent operators with a
single quark than with two different quarks.  The reason is that Fierz
identities relate operators, like for instance the counterpart of
${\cal O}^{(6)}_{qq'}$ with four equal quark fields, to the operators
${\cal O}_{i,q}^{(6)}$ with $i=1,\dots,4$. One way of implementing the
Fierz relations is to project Green's functions onto the basis that
includes so-called Fierz-evanescent operators, like ${\cal E}_7^{q}$
and ${\cal E}_8^{q}$ in Eq.~\eqref{eq:evan:2:LO} of
Appendix~\ref{App:unphysical}, that vanish due to Fierz identities.
SM operators with scalar or tensor currents do not contribute in our
calculation. This is most easily seen by inspecting their chiral and
Lorentz structure, neglecting operators with derivatives (see below).

Integrating out the $W$ and the $Z$ bosons at tree level gives the
following values for the Wilson coefficients at $\muew$
\begin{align}
\label{eq:initial:D1qq'}
{\cal D}_{1,qq'}^{(6)} & = 4 s_w^2 c_w^2 v_{q} v_{q'} 
                           + \frac{|I_q^3 - I_{q'}^3|}{6} |V_{qq'}|^2 \,,& 
{\cal D}_{2,qq'}^{(6)} & = 4 s_w^2 c_w^2 a_{q} a_{q'} 
    			   + \frac{|I_q^3 - I_{q'}^3|}{6} |V_{qq'}|^2 \,,& \\
\label{eq:initial:D3qq'}
{\cal D}_{3,qq'}^{(6)} & = 4 s_w^2 c_w^2 a_{q} v_{q'} 
                           - \frac{|I_q^3 - I_{q'}^3|}{6} |V_{qq'}|^2 \,,&
{\cal D}_{4,qq'}^{(6)} & =   {\cal D}_{5,qq'}^{(6)} 
                         = - {\cal D}_{6,qq'}^{(6)}  
			 =   |I_q^3 - I_{q'}^3| |V_{qq'}|^2 \,,&
\end{align}
and 
\begin{align}
\label{eq:initial:D1q}
&{\cal D}_{1,q}^{(6)} = 2 s_w^2 c_w^2 v_{q}^2 \,, &
&{\cal D}_{2,q}^{(6)} = 2 s_w^2 c_w^2 a_{q}^2 \,, & 
&{\cal D}_{3,q}^{(6)} = 4 s_w^2 c_w^2 v_{q} a_{q} \,,&
&{\cal D}_{4,q}^{(6)} = 0 \,.
\end{align}
Here, $s_w\equiv\sin \theta_w$, $c_w\equiv\cos \theta_w$, with
$\theta_w$ the weak mixing angle, while $I_q^3$ is the third component
of the weak isospin for the corresponding left-handed quark, i.e.,
$I_q^3 = 1/2$ for $q=u,c$ and $I_q^3 = -1/2$ for $q=d,s,b$. The CKM
matrix, $V_{qq'}$, will be set to unity unless specified otherwise,
while the vector and axial-vector couplings of the $Z$ boson to the
quarks are encoded in
\begin{align}
  &v_q \equiv  \frac{I_q^3 - 2 s_w^2 Q_q}{2 s_w c_w}\,,&
  &a_q \equiv -\frac{I_q^3}{2 s_w c_w}\,,&
\end{align}
where $Q_q$ is the electric charge of the corresponding quark.  Note
that ${\cal D}_{i,qq'}^{(6)} \equiv {\cal D}_{i,q'q}^{(6)}$ for
$i=1,2,4,5$, since the corresponding operators are symmetric under
$q\leftrightarrow q'$.

\section{Operator mixing and anomalous dimensions\label{sec:RGE}}
\begin{figure}
\includegraphics{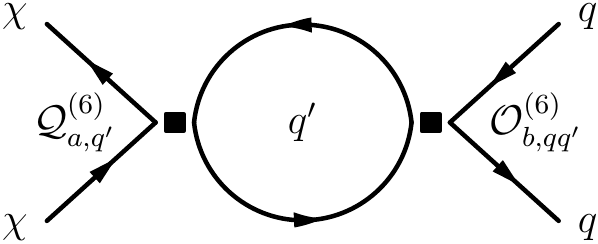}
\caption{A generic Feynman diagram with a double insertion of
  dimension-six operators, leading to the mixing into dimension-eight
  operators.
\label{fig:below_ew_fish}
}
\end{figure}

We are now ready to derive the leading contributions to the
DM--nucleon scattering rates for the case that, at the weak scale, DM
interacts with the visible sector only through the dimension-six
operators $\Q_{3,q}^{(6)}$ or $\Q_{4,q}^{(6)}$, with $q=b,c$. To
properly describe all the leading DM interactions we need to extend
the dimension-six effective Lagrangian ${\lag}_\chi$,
Eq.~\eqref{eq:d6:DM:Lf5}, to include the following dimension-eight
operators
\begin{equation}\label{eq:d8:Lf5}
{\lag}_\chi\supset - \frac{\sqrt{2} G_F}{\Lambda^2} \,
\sum_{\substack{q=u,d,s\\i=1,\ldots,4}} \C_{i,q}^{(8)} {\cal Q}_{i,q}^{(8)}\,,
\end{equation}
where
\begin{align}
\label{eq:dim8:Q12}
{\cal Q}_{1,q}^{(8)} & = \frac{m_q^2}{g_s^2} \, (\bar \chi \gamma_\mu \chi) (\bar q \gamma^\mu q)\,,
 &{\cal Q}_{2,q}^{(8)} &= \frac{m_q^2}{g_s^2} \, (\bar \chi\gamma_\mu\gamma_5 \chi)(\bar q \gamma^\mu q)\,, 
 \\ 
 \label{eq:dim8:Q34}
{\cal Q}_{3,q}^{(8)} & = \frac{m_q^2}{g_s^2} \, (\bar \chi \gamma_\mu \chi)(\bar q \gamma^\mu \gamma_5 q)\,,
& {\cal Q}_{4,q}^{(8)}& = \frac{m_q^2}{g_s^2} \, (\bar \chi\gamma_\mu\gamma_5 \chi)(\bar q \gamma^\mu \gamma_5 q)\,.
\end{align}
For future convenience, we defined the operators including two inverse
powers of the strong coupling constant.  Even if the Wilson
coefficient of the dimension-eight operators are zero at $\muew$, they
are generated below the electroweak scale from a double insertion of
one of the dimension-six operators in ${\lag}_\chi$ in
Eq.~\eqref{eq:d6:DM:Lf5} and one of the dimension-six operator from
${\lag}_\text{eff}^{\rm SM}$ in Eq.~\eqref{eq:Leff}, see
Fig.~\ref{fig:below_ew_fish}.\footnote{The only exception occurs when
  the values of the Wilson coefficients at the weak scale conspire to
  exactly cancel the divergence, so that the sum of the
  double-insertion diagrams is finite. This scenario is not fine tuned
  if it is protected by a symmetry.  A example of from the SM is the
  charm-quark contribution to the parameter $\epsilon_K$, where the
  GIM mechanism associated with the approximate flavor symmetry of the
  SM serves to cancel all divergences. We call the analogous mechanism
  for DM the {\it ``judicious operator equality GIM''}, in short
  ``Joe-GIM'' mechanism.  For Joe-GIM DM there is no mixing of
  dimension-six operators into dimension-eight operators below the
  weak scale~\cite{Witten:1976kx}. The leading contributions to the
  dimension-eight operators are then obtained by a finite one-loop
  matching calculation at the heavy-quark scales.}  The logarithmic
part of the running from $\muew$ to $\mustr$ gives
\begin{align}
\label{eq:C12:ewlog}
-\frac{\sqrt{2}G_F}{\Lambda^2}
\frac{m_q^2}{g_s^2} {\cal C}_{1(2),q}^{(8)}(\mustr)& 
\simeq 
\frac{\sqrt{2} G_F}{\Lambda^2}
\frac{12}{\pi^2} 
a_q s_w^2 c_w^2 
\sum_{q'=c,b} m_{q'}^2 v_{q'} {\cal
C}_{3(4),q'}^{(6)}(m_Z) \log \Big(\frac{m_{q'}}{m_Z}\Big),
\end{align}
where we set $\muew=m_Z$. This equation shows that the operators with
derivatives, for instance, $(\bar \chi \gamma_\mu \chi) \partial^2
(\bar q \gamma^\mu q)$, can be neglected because their effect on the
scattering rates is not enhanced by the large ratio
$m_{b,c}^2/m_q^2$. Furthermore, the set of operators in
Eqs.~\eqref{eq:dim8:Q12}--\eqref{eq:dim8:Q34} is closed under RG
running up to mass-dimension eight, if we keep only terms proportional
to two powers of the bottom- or charm-quark mass in the RG
evolution. At higher orders in QCD the purely electroweak expression
Eq.~\eqref{eq:C12:ewlog} gets corrected by terms of the order of
$\alpha_s^{n-1} \log^n(m_{b(c)}/m_Z)$. Since $m_{b(c)}\ll m_Z$, these
terms can amount to ${\mathcal O}(1)$ corrections. In the following we
resum these large QCD logarithms to leading-logarithmic order.

The techniques for the calculation of leading-logarithmic QCD
corrections with double insertions are standard~\cite{Witten:1976kx,
  Gilman:1982ap, Flynn:1989cf, Datta:1989xp, Herrlich:1994kh,
  Herrlich:1996vf, Brod:2013sga, Brod:2014qwa}. We first replace the
bare Wilson coefficients in
Eqs.~\eqref{eq:d6:DM:Lf5},~\eqref{eq:Leff}, and~\eqref{eq:d8:Lf5} with
their renormalized counterparts. The corresponding effective
Lagrangian reads
\begin{equation} \label{eq:Leff:ren}
\begin{split}
{\lag}_\text{eff} = & \frac{1}{\Lambda^2} \sum_{a} {\cal
  C}_a^{(6)} {\cal Q}_a^{(6)} - \sqrt{2} G_F \sum_{ab} {\cal D}_a^{(6)}
Z_{ab} {\cal O}_b^{(6)} \\ & - \frac{\sqrt{2} G_F}{\Lambda^2} \bigg \{
\sum_{ab} {\cal C}_a^{(8)} \tilde Z_{ab} {\cal Q}_b^{(8)} + \sum_{abc}
    {\cal D}_a^{(6)} {\cal C}_b^{(6)} \hat Z_{ab,c} {\cal Q}_c^{(8)}
    \bigg \} \,.
\end{split}
\end{equation}
The compound indices $a$, $b$, $c$, run over both the operator labels
and quark-flavor indices.  In Eq.~\eqref{eq:Leff:ren}, we have already
made use of the fact that the QCD anomalous dimensions of the
operators $\Q_a^{(6)}$ in
Eqs.~\eqref{eq:dim6:Q1Q2}--\eqref{eq:dim6:Q3Q4} are zero, and have not
introduced the corresponding renormalization constants.

In dimension regularization around $d = 4 - 2\epsilon$ space-time dimensions,
the renormalization constants admit a double expansion in the strong 
coupling constant and $\epsilon$
\begin{equation}
Z_{ab} = \delta_{ab} + \frac{\alpha_s}{4\pi} \sum_{k=0,1}
\frac{1}{\epsilon^k} Z_{ab}^{(1,k)} + {\cal O}(\alpha_s^2) \,,
\end{equation}
and similarly for $\tilde Z$ and $\hat Z$.

The RG evolution of the dimension-six Wilson coefficients is
determined by a RG equation that is linear in the Wilson coefficients,
\begin{equation}\label{eq:d6:rge}
\mu\frac{d}{d\mu} {\cal D}_{a}^{(6)} (\mu) = \sum_b {\cal D}_{b}^{(6)}
(\mu) \gamma_{ba}\,,
\qquad\text{with}\quad \gamma=-\frac{d \log Z}{d \log \mu}\,.
\end{equation}
On the other hand, the running of the dimension-eight Wilson
coefficients receives two contributions.  In addition to the running
of the $m_q^2/g_s^2$ prefactor, encoded in $\tilde \gamma$, there are
contributions from double insertions of dimension-six operators, see
Fig.~\ref{fig:below_ew_fish}. This leads to a RG equation that is
quadratic in dimension-six Wilson coefficients~\cite{Herrlich:1994kh,
  Herrlich:1996vf},
\begin{equation}\label{eq:d8:rge}
\mu\frac{d}{d\mu} {\cal C}_{a}^{(8)} (\mu) = \sum_b {\cal C}_{b}^{(8)}
(\mu) \tilde \gamma_{ba} + \sum_{bc} {\cal D}_{b}^{(6)} (\mu) {\cal
  C}_{c}^{(6)} (\mu) \hat\gamma_{bc,a} \,.
\end{equation}
To leading order in the strong coupling constant the rank-three
anomalous dimension tensor $\hat \gamma_{ab,c}$~\cite{Herrlich:1994kh,
  Herrlich:1996vf} is given by
\begin{equation}
\hat\gamma_{ab,c} = \frac{\alpha_s}{2\pi} \hat{Z}_{ab,c}^{(1,1)} +
            {\cal O}(\alpha_s^2) \,.
\end{equation}
Next we provide the explicit values for the anomalous dimensions.  In
our notation, the anomalous dimensions are expanded in powers of
$\alpha_s$,
\begin{equation}
\gamma=\gamma^{(1)}+\gamma^{(2)}+\cdots\,,
\end{equation}
with $\gamma^{(n)}\propto (\alpha_s/4\pi)^n$, and similarly for
$\tilde \gamma$ and $\hat \gamma$.  

We start by giving the results for the mixing of the dimension-eight
operators coupling DM to quarks, Eqs.~\eqref{eq:dim8:Q12}
and~\eqref{eq:dim8:Q34}. This mixing is encoded in the $\tilde \gamma$
and $\hat \gamma$ anomalous dimensions. We obtain the $\hat \gamma$
from the poles of the double insertions, Fig.~\ref{fig:below_ew_fish}.
The only nonzero entries leading to mixing into operators with
light-quark currents are
\begin{equation}\label{eq:gammahat:val}
\hat \gamma_{{\cal O}_{3,q'q}^{(6)}, {\cal Q}_{3,q'}^{(6)}; {\cal Q}_{1,q}^{(8)}}^{(1)} = 
\hat \gamma_{{\cal O}_{3,q'q}^{(6)}, {\cal Q}_{4,q'}^{(6)}; {\cal Q}_{2,q}^{(8)}}^{(1)} = 
\hat \gamma_{{\cal O}_{2,qq'}^{(6)}, {\cal Q}_{3,q'}^{(6)}; {\cal Q}_{3,q}^{(8)}}^{(1)} = 
\hat \gamma_{{\cal O}_{2,qq'}^{(6)}, {\cal Q}_{4,q'}^{(6)}; {\cal Q}_{4,q}^{(8)}}^{(1)} = 
- 16 N_c \frac{m_{q'}^2}{m_{q}^2} \frac{\alpha_s}{4\pi} \,.
\end{equation}
The remaining contribution to the RG running of the dimension-eight
operators is entirely due to the $m_q^2/g_s^2$ prefactors in the
definition of the operators, namely
\begin{equation}
\tilde \gamma_{ab} = 2(\gamma_m^{(0)} - \beta_0) \frac{\alpha_s}{4\pi} \delta_{ab} \,,
\quad\text{where}\quad
\gamma_m^{(0)} = 6 C_F\quad\text{and}\quad \beta_0 = 11 - \frac{2}{3} N_f \,,
\end{equation}
with $C_F=4/3$ for QCD, and $N_f$ the number of active quark flavors.

\begin{figure}[]
\centering
\includegraphics{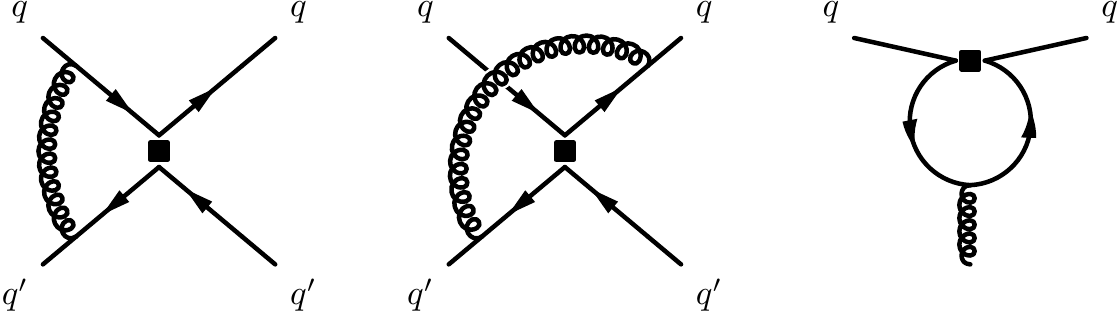}
\caption{Selection of diagrams entering the computation of
QCD one-loop anomalous dimension of four-fermion operators. 
The poles of QCD penguin diagrams affect the mixing
via the e.o.m.-vanishing operators defined in 
Appendix~\ref{App:unphysical}.
\label{fig:4fermionrunning}}
\end{figure}

The RG running of the dimension-six operators in the SM weak effective
Lagrangian is due to one-loop gluon exchange diagrams, see
Fig.~\ref{fig:4fermionrunning}.  Since the corresponding anomalous
dimension matrix $\gamma$ has many entries, we split the result into
several blocks.

The anomalous dimension matrix in the subsector spanned by the
operators in Eqs.~\eqref{eq:op:q:12}--\eqref{eq:op:q:34},
\begin{equation}
\big({\cal O}_{1,q}^{(6)},\,{\cal O}_{2,q}^{(6)},\,{\cal O}_{3,q}^{(6)},\,{\cal O}_{4,q}^{(6)}\big)\,, 
\end{equation} 
reads
\begin{equation}
\gamma^{(1)} = \frac{\alpha_s}{4\pi}
\begin{pmatrix}
 4&4& 0& - \frac{28}{3}\\
 0& 0& 0&\frac{44}{3}\\
 0& 0&\frac{44}{9}& 0\\
 \frac{5}{3}&\frac{13}{3}& 0& - \frac{106}{9}
\end{pmatrix}.
\label{eq:ADMdiagonal2diagonal}
\end{equation}
Note that, at one-loop, there is no mixing into operators with a
different quark flavor.

The anomalous dimensions describing the mixing of the same operators,
\begin{equation}
\big({\cal O}_{1,q}^{(6)},\,{\cal O}_{2,q}^{(6)},\,{\cal O}_{3,q}^{(6)},\,{\cal O}_{4,q}^{(6)}\big)\,, 
\end{equation} 
into the operators
\begin{equation}
\big({\cal O}_{1,qq'}^{(6)},\,{\cal O}_{2,qq'}^{(6)},\,{\cal O}_{3,qq'}^{(6)},\,{\cal O}_{3,q'q}^{(6)},\,{\cal O}_{4,qq'}^{(6)},\,{\cal O}_{5,qq'}^{(6)},\,{\cal O}_{6,qq'}^{(6)},\,{\cal O}_{6,q'q}^{(6)}\big)\,,
\end{equation} 
read
\begin{equation}
\gamma^{(1)} = \frac{\alpha_s}{4\pi}
\begin{pmatrix}
 0& 0& 0& 0&\frac{8}{3}& 0& 0& 0\\
 0& 0& 0& 0&\frac{8}{3}& 0& 0& 0\\
 0& 0& 0& 0& 0& 0& \frac{8}{3}& 0 \\
 0& 0& 0& 0&\frac{20}{9}& 0& 0& 0
\end{pmatrix}.
\end{equation}

The anomalous dimension matrix in the subsector spanned by the
operators in Eqs.~\eqref{eq:op:qq':12}--\eqref{eq:op:qq':56},
\begin{equation}
\big({\cal  O}_{1,qq'}^{(6)},\,{\cal O}_{2,qq'}^{(6)},\,{\cal O}_{3,qq'}^{(6)},\,{\cal O}_{3,q'q}^{(6)},\,{\cal O}_{4,qq'}^{(6)},\,{\cal O}_{5,qq'}^{(6)},\,{\cal O}_{6,qq'}^{(6)},\,{\cal O}_{6,q'q}^{(6)} \big)\,,
\end{equation} 
reads 
\begin{equation}
\gamma^{(1)} = \frac{\alpha_s}{4\pi}
\begin{pmatrix}
 0& 0& 0& 0& 0&12& 0& 0\\
 0& 0& 0& 0&12& 0& 0& 0\\
 0& 0& 0& 0& 0& 0& 0&12\\
 0& 0& 0& 0& 0& 0&12& 0\\
 0&\frac{8}{3}& 0& 0& - \frac{19}{3}&5& 0& 0\\
\frac{8}{3}& 0& 0& 0&5& - 9& 0& 0\\
 0& 0& 0&\frac{8}{3}& 0& 0& - \frac{23}{3}&5\\
 0& 0&\frac{8}{3}& 0& 0& 0&5& - \frac{23}{3}
\end{pmatrix}.
\end{equation}

The part of the anomalous dimension matrix mixing the operators, 
\begin{equation}
\big({\cal  O}_{1,qq'}^{(6)},\,{\cal O}_{2,qq'}^{(6)},\,{\cal O}_{3,qq'}^{(6)},\,{\cal O}_{3,q'q}^{(6)},\,{\cal O}_{4,qq'}^{(6)},\,{\cal O}_{5,qq'}^{(6)},\,{\cal O}_{6,qq'}^{(6)},\,{\cal O}_{6,q'q}^{(6)}\big)\,,
\end{equation} 
into the same operators, but with different quark flavor structure,
$q''\ne q'$,
\begin{equation}
\big({\cal O}_{1,qq''}^{(6)},\,{\cal O}_{2,qq''}^{(6)},\,{\cal O}_{3,qq''}^{(6)},\,{\cal O}_{3,q''q}^{(6)},\,{\cal O}_{4,qq''}^{(6)},\,{\cal O}_{5,qq''}^{(6)},\,{\cal O}_{6,qq''}^{(6)},\,{\cal O}_{6,q''q}^{(6)}\big)\,,
\end{equation} 
is 
\begin{equation}
\gamma^{(1)} = \frac{\alpha_s}{4\pi}
\diag\Big(
 0, 0, 0, 0, \frac{4}{3}, 0, \frac{4}{3}, 0\Big).
 \end{equation}
    	  
Finally, the part of the anomalous dimension matrix mixing the operators
\begin{equation}
\big({\cal
  O}_{1,qq'}^{(6)}, {\cal O}_{2,qq'}^{(6)}, {\cal O}_{3,qq'}^{(6)},
{\cal O}_{3,q'q}^{(6)}, {\cal O}_{4,qq'}^{(6)}, {\cal
  O}_{5,qq'}^{(6)}, {\cal O}_{6,qq'}^{(6)}, {\cal O}_{6,q'q}^{(6)}\big)
\end{equation} 
into the operators
\begin{equation}
\big({\cal O}_{1,q}^{(6)}, {\cal O}_{2,q}^{(6)}, {\cal
  O}_{3,q}^{(6)}, {\cal O}_{4,q}^{(6)}\big)
\end{equation} 
has only two nonzero entries,
\begin{equation}
\gamma^{(1)} = \frac{\alpha_s}{4\pi}
\begin{pmatrix}
 0& 0& 0& 0\\
 0& 0& 0& 0\\
 0& 0& 0& 0\\
 0& 0& 0& 0\\
 0& 0& 0&\frac{4}{3}\\
 0& 0& 0& 0\\
 0& 0&\frac{4}{9}& 0\\
 0& 0& 0& 0
\end{pmatrix}.
\end{equation}

All the remaining entries in $\gamma$ vanish.  We extracted the
anomalous dimensions from the off-shell renormalization of Green's
functions with appropriate external states. We checked explicitly that
our results are gauge-parameter independent.  In
Appendix~\ref{App:unphysical}, we list the evanescent 
and e.o.m.-vanishing operators that enter at intermediate stages 
of the computation.

\section{Renormalization Group evolution\label{sec:match}}

With the anomalous dimensions of Section~\ref{sec:RGE}, we can now
compute the Wilson coefficients at $\mustr\sim 2$~GeV in terms of
initial conditions at the weak scale. The Wilson coefficients ${\cal
  C}_{b}^{(6)}$ do not run, thus ${\cal C}_{b}^{(6)}(\mustr)={\cal
  C}_{b}^{(6)}(\muew)$.  The RG running for the remaining Wilson
coefficients is controlled by the RG equations in
Eqs.~\eqref{eq:d6:rge} and \eqref{eq:d8:rge}, which we combine into a
single expression,
\begin{equation}\label{eq:rge:eff}
\mu\frac{d}{d\mu} \mathscr{C}_{a} (\mu) = \sum_b \mathscr{C}_{b} (\mu)
\gamma_{ba}^\text{eff} \,.
\end{equation}
Here, we defined a vector of Wilson coefficients as
\begin{equation}\label{eq:D:eff}
\mathscr{C} (\mu) \equiv 
\begin{pmatrix}
{\cal D}^{(6)} (\mu)
\\
{\cal C}^{(8)} (\mu)
\end{pmatrix}.
\end{equation}
and absorbed the (scale-independent) Wilson coefficients
$\C_a^{(6)}$ into the effective anomalous-dimension matrix
\begin{equation}\label{eq:gamma:D:eff}
\gamma^\text{eff} \equiv 
\begin{pmatrix}
\gamma&{\cal C}^{(6)} \cdot \hat\gamma
\\
0&\tilde \gamma
\end{pmatrix}\qquad
\text{with}
\qquad
({\cal C}^{(6)} \cdot \hat\gamma )_{ba} \equiv \sum_{c} 
{\cal  C}_{c}^{(6)} \hat\gamma_{bc,a}\,.
\end{equation}
Since the $\C_a^{(6)}$ Wilson coefficients are RG invariant, the
tensor product effectively transforms the rank-three tensor $\hat
\gamma_{ab,c}$ into an equivalent matrix, ${\cal C}^{(6)} \cdot
\hat\gamma$, with all its entries constant, that is equivalent to the
tensor for the purpose of RG running.  This has the advantage that one
can use the standard methods for single insertions to solve the RG
equations.

The RG evolution proceeds in multiple steps.  The first step is the
matching of the (complete or effective) theory of DM interactions
above the weak scale onto the five-flavor EFT.  This matching
computation yields the initial conditions for $\C_a^{(6)}(\muew)$ and
$\C_a^{(8)}(\muew)$ at the weak scale.  At leading-logarithmic order
it suffices to perform the matching at $\muew$ at tree-level. If the
mediators have weak-scale masses, we obtain $\C_a^{(6)}(\muew)\ne 0$
and $\C_a^{(8)}(\muew)=0$.  If the mediators are much heavier than the
weak scale, with masses of order $M \gg m_Z$, the RG running above the
electroweak scale can induce nonzero $\C_a^{(8)}(\muew) \sim
\log(\muew/M)$. We discuss the latter case in Section \ref{sec:UV}.
For the RG evolution below the electroweak scale one also needs the
coefficients ${\cal D}_a^{(6)}(\muew)$. The SM contributions to the
tree-level initial conditions for ${\cal D}_a^{(6)}(\muew)$ are
provided in Eqs.~\eqref{eq:initial:D1qq'}--\eqref{eq:initial:D1q}.

The second step is to evolve $\C_a^{(8)}(\mu)$ and ${\cal
  D}_a^{(6)}(\mu)$ from the electroweak scale to lower scales
according to Eq.~\eqref{eq:rge:eff}. The RG evolution is in a theory
with $N_f=5$ quark flavors, when $\mu_b\leq \mu \leq \muew$, with
$N_f=4$, when $\mu_c\leq \mu \leq \mu_b$, and with $N_f=3$, when
$\mustr\leq \mu \leq \mu_c$.  Here the $\mu_{b(c)}$ denote the
threshold scale at which the bottom(charm)-quark is removed from the
theory.  In our numerical analysis we will use $\mu_b=4.18$\,GeV and
$\mu_c=2$~GeV.  At leading-logarithmic order, there are no non-trivial
matching corrections at the bottom- and charm-quark thresholds, and we
simply have
\begin{equation}
\label{eq:matching:heavyquark}
{\cal C}_{a}^{(d)}(\mu_b)\big|_{N_f=4}= 
{\cal C}_{a}^{(d)}(\mu_b)\big|_{N_f=5}\,, 
\qquad\quad 
{\cal C}_{a}^{(d)}(\mu_c)\big|_{N_f=3}= 
{\cal C}_{a}^{(d)}(\mu_c)\big|_{N_f=4}\,.
\end{equation}
This means that we can switch to the EFT with four active quark
flavors by simply dropping all operators in Eq.~\eqref{eq:Leff:ren}
that involve a bottom-quark field, and to the EFT with three active
quark flavors by simply dropping all operators with charm-quark
fields.  The leading-order matching at $\mu_{q'} \sim m_{q'}$ comes
with a small uncertainty due to the choice of matching scale that is
of order $\log(\mu_{q'} / m_{q'})$.  This is formally of higher order
in the RG-improved perturbation theory. The uncertainty is canceled in
a calculation at next-to-leading-logarithmic order by finite threshold
corrections at the respective threshold scale.

\bigskip

This is a good point to pause and compare our results with the
literature.  The RG evolution of the operators in
Eqs.~\eqref{eq:dim6:Q1Q2}--\eqref{eq:dim6:Q3Q4} below the electroweak
scale has been studied in Ref.~\cite{DEramo:2014nmf}, which
effectively resummed the large logarithms $\log (\mustr / \muew)$ to
all orders in the {\it Yukawa} couplings.  Such a resummation is
problematic for two reasons.  Firstly, it does not take into account
that the RG evolution stops at the heavy-quark thresholds, below which
the Wilson coefficients are RG invariant.  (Below the heavy-quark
thresholds, there are no double insertions with heavy quarks and the
running of the factor $m_q^2/g_s^2$ is precisely canceled by the
running of the Wilson coefficients of the dimension-eight operators.)
This introduces a spurious scale dependence of the Wilson coefficients
in the three-flavor EFT, of the order of $|\log(\mustr / m_{b(c)})|
\lesssim 50\%$, that is not canceled by the hadronic matrix elements.
Secondly, such a resummation is not consistent within the EFT
framework.  Since there are no Higgs-boson exchanges in the EFT below
the weak scale, the scheme-dependence of the anomalous dimensions and
the residual matching scale dependence at the heavy-quark thresholds
is not consistently canceled by higher-orders, leading to unphysical
results.

\bigskip

Continuing with our analysis, the final step is to match the
three-flavor EFT onto the EFT with nonrelativistic neutrons and
protons that is then used to predict the scattering rates for DM on
nuclei using nuclear response functions.  The matching for the
dimension-eight contributions proceeds in exactly the same way as
described in Refs.~\cite{Bishara:2016hek, Bishara:2017pfq} for the
operators up to dimension seven.  In practice, this means that we
obtain the following contributions to the nonrelativistic coefficients
(see Refs.~\cite{Fitzpatrick:2012ix, Anand:2013yka, Bishara:2016hek,
  Bishara:2017pfq, Bishara:2017nnn}),
\begin{align}
\label{eq:cNR1}
c_{1}^p &= - \frac{1}{4\pi \alpha_s} \frac{\sqrt{2} G_F}{\Lambda^2} \sum_{q=u,d,s} m_q^2 
\C_{1,q}^{(8)}\, 
F_{1}^{q/p}
+\cdots \,, 
\\
\label{eq:cNR4}
c_{4}^p &= ~\,\phantom{+}\frac{1}{\pi\alpha_s}\frac{\sqrt{2} G_F}{\Lambda^2}\sum_{q=u,d,s} m_q^2 
\C_{4,q}^{(8)}\, 
F_{A}^{q/p} 
+\cdots\,,
\\
\label{eq:cNR6}
c_{6}^p &= - \frac{1}{4\pi\alpha_s} \frac{\sqrt{2} G_F}{\Lambda^2}\sum_{q=u,d,s} m_q^2 
\C_{4,q}^{(8)} \,
F_{P'}^{q/p}
+\cdots\,,
\\
\label{eq:cNR7}
c_{7}^p &= \phantom{+} \frac{1}{2\pi\alpha_s} \frac{\sqrt{2} G_F}{\Lambda^2} \sum_{q=u,d,s} m_q^2 
\C_{3,q}^{(8)} \,
F_{A}^{q/p}
+\cdots \,,
\\
\label{eq:cNR8}
c_{8}^p &= - \frac{1}{2\pi\alpha_s} \frac{\sqrt{2} G_F}{\Lambda^2} \sum_{q=u,d,s} m_q^2 
\C_{2,q}^{(8)} \,
F_{1}^{q/p} 
+\cdots\,,
\\
\label{eq:cNR9}
c_{9}^p &= - \frac{1}{2\pi\alpha_s}\frac{\sqrt{2} G_F}{\Lambda^2} \sum_{q=u,d,s} m_q^2 \Big[ 
	\C_{2,q}^{(8)} \big( F_{1}^{q/p} + F_{2}^{q/p}\big)  
	+ \frac{m_p}{m_\chi}\,\C_{3,q}^{(8)}\, F_{A}^{q/p}\Big] +\cdots \,,
\end{align}
and similarly for neutrons, with $p\to n$. The quark masses and the
strong-coupling constant in these expressions should be evaluated in
the three-flavor theory at the same scale as the nuclear response
functions, i.e., $\mustr = 2\,$GeV.  The ellipses denote the
contributions from dimension-six interactions proportional to
$\C_a^{(6)}$ as well as the contributions due to dimension-five and
dimension-seven operators, which can be found in Eqs.~(17)--(24) of
Ref.~\cite{Bishara:2017pfq}.

The strong coupling $\alpha_s$ appears in
Eqs. \eqref{eq:cNR1}--\eqref{eq:cNR9} as a consequence of the
$1/g_s^2$ prefactor in the definition of the dimension-eight operators
in Eqs.~\eqref{eq:dim8:Q12}--\eqref{eq:dim8:Q34}. When expanding the
resummed results in the strong coupling constant, the $\alpha_s$
cancels in the leading expressions, and we find
\begin{align}
\label{eq:cNR1:exp}
c_{1}^p &\simeq \phantom{+}\frac{3}{\pi^2} \frac{\sqrt{2} G_F}{\Lambda^2} \sum_{q'=c,b} m_{q'}^2 \, 
\log({\mu_{q'}}/{m_Z})\,
\C_{3,q'}^{(6)}(m_Z)\,
\sum_{q=u,d,s} {\cal D}_{3,q'q}\,F_{1}^{q/p} +{\mathcal O}(\alpha_s)+\ldots \,, 
\\
\label{eq:cNR4:exp}
c_{4}^p &\simeq - \frac{12}{\pi^2}\frac{\sqrt{2} G_F}{\Lambda^2} \sum_{q'=c,b} m_{q'}^2  \, 
\log({\mu_{q'}}/{m_Z})\,
\C_{4,q'}^{(6)}(m_Z)\,
\sum_{q=u,d,s} {\cal D}_{2,qq'}\,F_{A}^{q/p}  +{\mathcal O}(\alpha_s) +\ldots\,,
\\
\label{eq:cNR6:exp}
c_{6}^p &\simeq \phantom{+}\frac{3}{\pi^2}\frac{\sqrt{2} G_F}{\Lambda^2} \sum_{q'=c,b} m_{q'}^2  \, 
\log({\mu_{q'}}/{m_Z})\,
\C_{4,q'}^{(6)}(m_Z)\,
\sum_{q=u,d,s} {\cal D}_{2,qq'}\,F_{P'}^{q/p} +{\mathcal O}(\alpha_s)+\ldots\,,
\\
\label{eq:cNR7:exp}
c_{7}^p &\simeq - \frac{6}{\pi^2}\frac{\sqrt{2} G_F}{\Lambda^2} \sum_{q'=c,b} m_{q'}^2  \, 
\log({\mu_{q'}}/{m_Z})\,
\C_{3,q'}^{(6)}(m_Z)\,
\sum_{q=u,d,s} {\cal D}_{2,qq'}\,F_{A}^{q/p}  +{\mathcal O}(\alpha_s)+\ldots \,,
\\
\label{eq:cNR8:exp}
c_{8}^p &= \phantom{+}\frac{6}{\pi^2} \frac{\sqrt{2} G_F}{\Lambda^2} \sum_{q'=c,b} m_{q'}^2  \, 
\log({\mu_{q'}}/{m_Z})\,
\C_{4,q'}^{(6)}(m_Z)\,
\sum_{q=u,d,s} {\cal D}_{3,q'q}\,F_{1}^{q/p}  +{\mathcal O}(\alpha_s)+\ldots \,,
\\
\begin{split}
\label{eq:cNR9:exp}
c_{9}^p &= \phantom{+}\frac{6}{\pi^2}\frac{\sqrt{2} G_F}{\Lambda^2} \sum_{q'=c,b} m_{q'}^2  \,  
\log({\mu_{q'}}/{m_Z})\,
\sum_{q=u,d,s}  \Big\{ \C_{4,q'}^{(6)}(m_Z)\, {\cal D}_{3,q'q}\, \big( F_{1}^{q/p} + F_{2}^{q/p}
\big)   
\\
&\hspace{16.2em}
+ \frac{m_N}{m_\chi} \C_{3,q'}^{(6)}(m_Z)\, {\cal D}_{2,qq'}\, F_{A}^{q/p}    \Big\} +{\mathcal O}(\alpha_s)+\ldots \,.
\end{split}
\end{align}
The quark masses in these expressions should be evaluated at the weak
scale, $m_{q'} = m_{q'}(m_Z)$, while $\mu_{q'}$ is the scale at which
the $q'$ quark is integrated out.  We have provided the SM Wilson
coefficients, ${\cal D}_{2,qq'}$ and ${\cal D}_{3,q'q}$, in
Eqs.~\eqref{eq:initial:D1qq'} and \eqref{eq:initial:D3qq'}.  The
expanded equations clearly illustrate that the leading terms are of
electroweak origin, and thus of ${\mathcal O}(\alpha_s^0)$, while the
corrections due to QCD resummation start at ${\mathcal O}(\alpha_s)$.

\subsection{Numerical analysis and the impact of resummation}

\begin{figure}[t]
\centering
\includegraphics{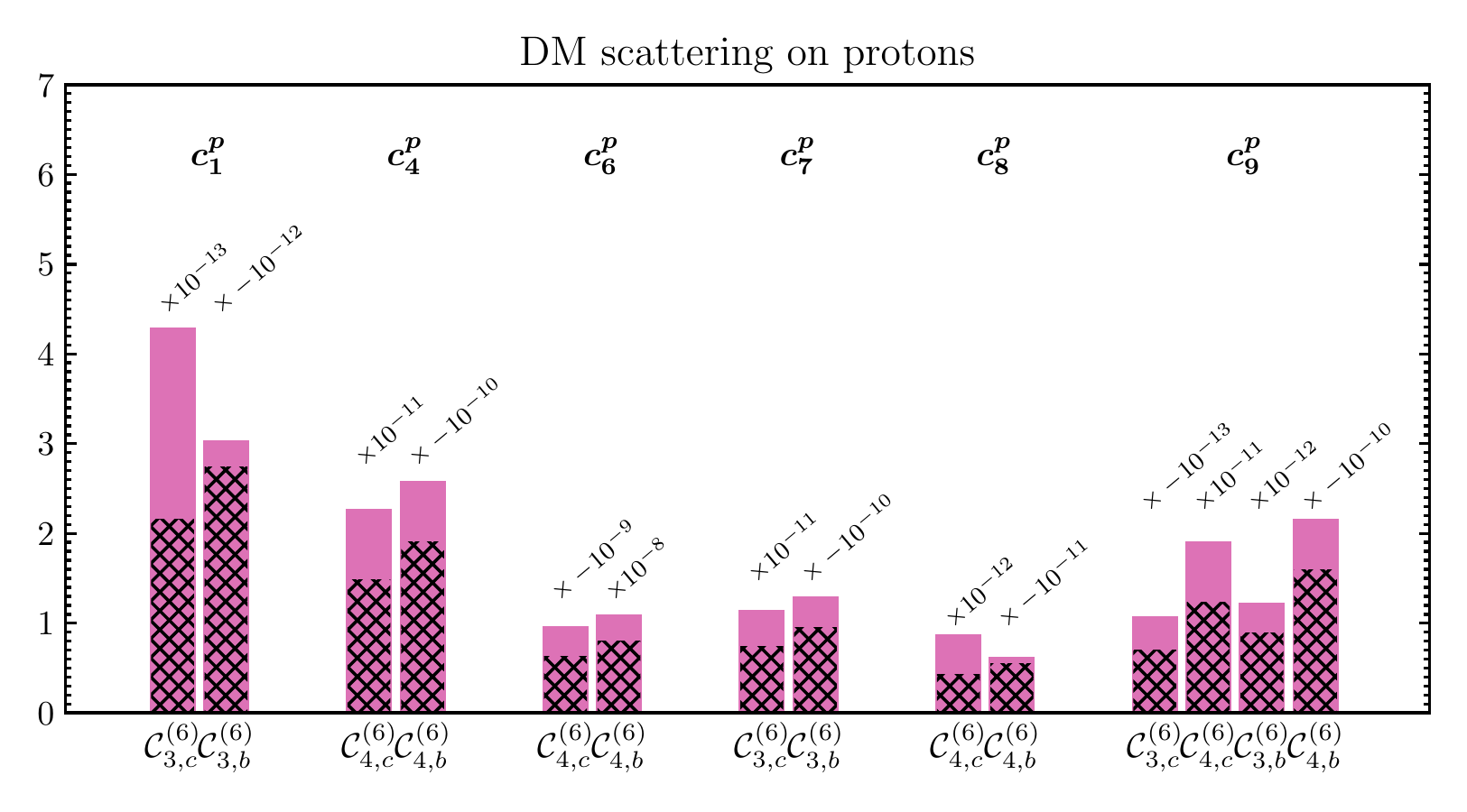}
\caption{In magenta: the values of nonrelativistic coefficients
  $c^p_i$, Eqs.~\eqref{eq:cNR1}--\eqref{eq:cNR9}, controlling the
  scattering of DM on protons, taking only one of the Wilson
  coefficients nonzero, setting it to ${\cal
    C}^{(d)}_{a}=1/\Lambda^2$, with $\Lambda=1$\,TeV.  Hatched: the
  results without QCD resummation,
  Eqs.~\eqref{eq:cNR1:exp}--\eqref{eq:cNR9:exp}.\label{fig:cpi}}
\end{figure}

\begin{figure}[t]
\centering
\includegraphics{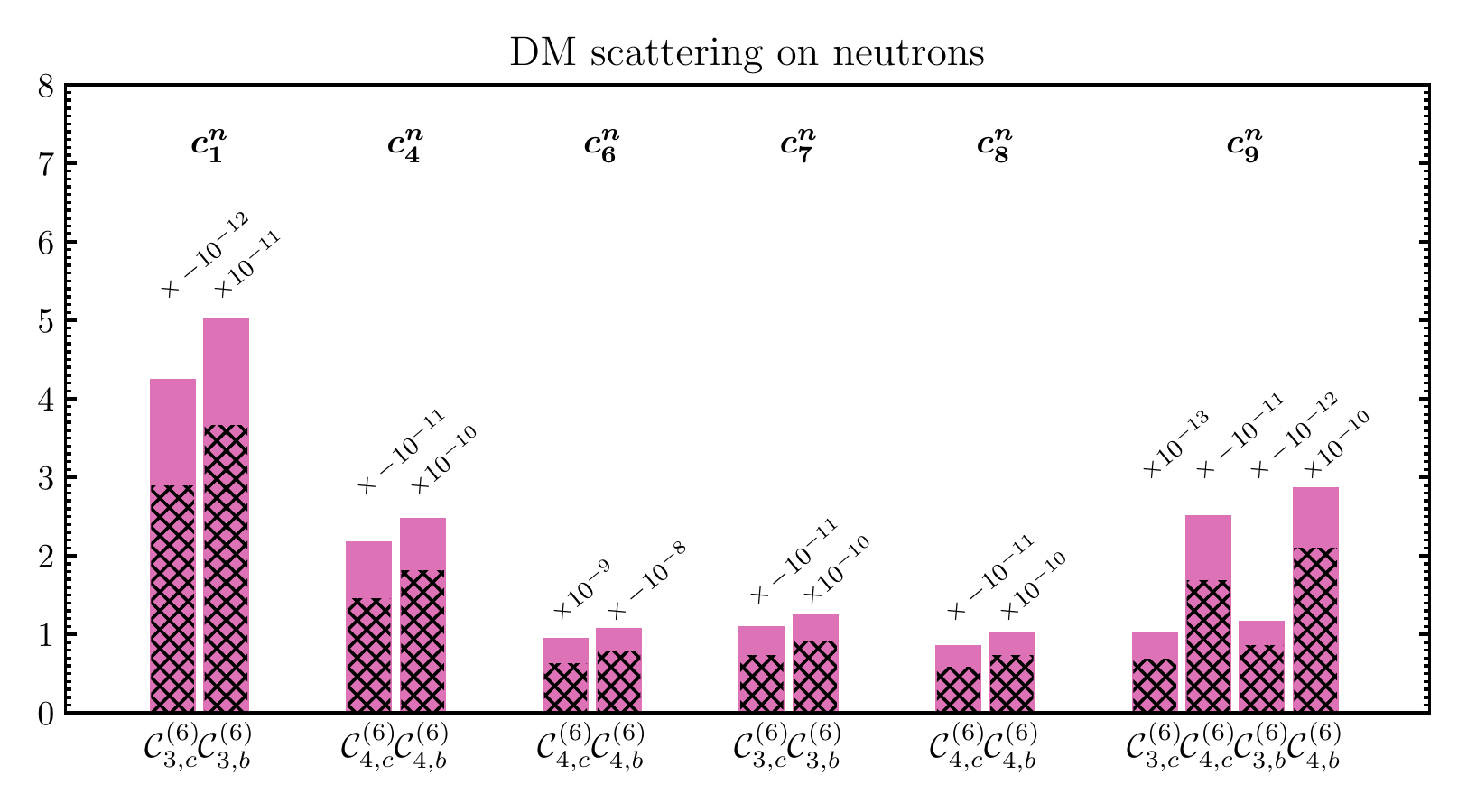}
\caption{The same as in Fig. \ref{fig:cpi} but for DM couplings to neutrons, $c^n_i$.
\label{fig:cni}}
\end{figure}

In Figs.~\ref{fig:cpi} and~\ref{fig:cni} we show two numerical
examples that illustrate the relative importance of the above results.
We set $\Lambda=1$\,TeV and switch on a single dimension-six Wilson
coefficient, ${\cal C}^{(d)}_{3,c}$, ${\cal C}^{(d)}_{4,c}$, ${\cal
  C}^{(d)}_{3,b}$, or ${\cal C}^{(d)}_{4,b}$ at a time, setting it to
${\cal C}^{(d)}_{a}=1$.  Fig.~\ref{fig:cpi} shows the resulting
nonrelativistic couplings for SM scattering on protons, $c_i^p$.  The
magenta columns are the full results, including QCD resummation.  The
hatched columns give the results without the QCD resummation from
Eqs.~\eqref{eq:cNR1:exp}--\eqref{eq:cNR9:exp}.  Fig.~\ref{fig:cni}
shows the corresponding results for DM couplings to neutrons, $c_i^n$.

In these examples we set $\mustr=\mu_c=2$ GeV and used the following
quark masses at $\mu = m_Z$,
\begin{align*}
m_u (m_Z) & = 1.4  \, \text{MeV} \,, & 
m_d (m_Z) & = 3.1  \, \text{MeV} \,, &
m_s (m_Z) & = 63   \, \text{MeV} \,, &\\
m_c (m_Z) & = 0.78 \, \text{GeV} \,, &
m_b (m_Z) & = 3.1  \, \text{GeV} \,. &
\end{align*}
These were obtained using the one-loop QCD running to evolve the
$\overline{\rm MS}$ quark masses $m_{u,d,s}(2\,\text{GeV})$ and
$m_{c(b)}(m_{c(b)})$, taken from Ref.~\cite{Patrignani:2016xqp}, to
the common scale $\mu=m_Z$.  For the nuclear coefficients that depend
on the DM mass and/or the momentum transfer, we choose $m_\chi =
100\,$GeV and a momentum transfer of $q = 50\,$MeV.

As seen from Figs.~\ref{fig:cpi} and~\ref{fig:cni}, the resummation of
QCD logarithms enhances $c^{p(n)}_i$ by approximately $10\%$ to $50\%$
depending on the specific case.  The typical enhancement is ${\cal
  O}(30\%)$. In the numerics we have set the CKM matrix element to
unity, thus ignoring all flavor changing transitions. This is a very
good approximation for operators with bottom quarks. For charm quarks
the effect of flavor off-diagonal CKM matrix elements is more
important, yet still subleading.  If we including the off-diagonal
terms in Eqs.~\eqref{eq:initial:D1qq'}--\eqref{eq:initial:D3qq'}, then
the largest correction reaches $16\%$ for $c_1^{p}$ induced from
${\cal C}^{(6)}_{3,c}$ ($31\%$ for the result without resummation), as
there is an up to $10\%$ cancellation between the ${\cal D}_{3,cu}$
and ${\cal D}_{3,cd}$ contributions with respect to the case of unit
CKM matrix. For all other cases the error due to setting the CKM
matrix to unity is less than $10\%$.

Finally, we compare the contributions to DM scattering originating
from electroweak corrections as opposed to the intrinsic charm and
bottom axial charges. For the case of axial-vector--axial-vector
interactions ($\C_{4,c}^{(6)} \neq 0$, $\C_{4,b}^{(6)} \neq 0$) we
have
\begin{align}
&\text{weak:}      & c_4^p &\simeq \phantom{+}\frac{1}{\Lambda^2}\Big(0.02\,  \C_{4,c}^{(6)}-0.26\,  \C_{4,b}^{(6)}\Big) \cdot 10^{-3},
\\
&\text{intrinsic:} & c_4^p &\simeq          -\frac{4}{\Lambda^2} \Big(\Delta c\, \C_{4,c}^{(6)}+\Delta b\, \C_{4,b}^{(6)} \Big)\sim  -\frac{1}{\Lambda^2} \Big(2\, \C_{4,c}^{(6)}+0.2  \C_{4,b}^{(6)}\Big) \cdot 10^{-3}.
\end{align}
We see that for the bottom quarks the weak contribution,
Eq.~\eqref{eq:cNR4:exp}, is comparable to the contribution from the
intrinsic bottom axial charge, while for charm quarks the contribution
due to the intrinsic charm axial charge dominates.

For vector--axial-vector interactions ($\C_{3,c}^{(6)} \neq 0$,
$\C_{3,b}^{(6)} \neq 0$) we have
\begin{align}
&\text{weak:}&c_1^p&\simeq\phantom{+} \frac{1}{\Lambda^2}\Big(0.4\,  \C_{3,c}^{(6)}-3.0\,  \C_{3,b}^{(6)}\Big) \cdot 10^{-6},
\\
&\text{intrinsic:}&c_7^p&\simeq -\frac{2}{\Lambda^2} \Big(\Delta c\, \C_{3,c}^{(6)}+\Delta b\, \C_{3,b}^{(6)} \Big)\sim  -\frac{1}{\Lambda^2} \Big( \C_{3,c}^{(6)}+0.1  \C_{3,b}^{(6)}\Big) \cdot 10^{-3},
\\
&&c_9^p&\simeq \frac{2}{\Lambda^2} \frac{m_N}{m_\chi}\Big(\Delta c\, \C_{3,c}^{(6)}+\Delta b\, \C_{3,b}^{(6)} \Big)\sim  -\frac{1}{\Lambda^2} \Big(9\, \C_{3,c}^{(6)}+0.9  \C_{3,b}^{(6)}\Big) \cdot 10^{-6},
\end{align}
where in the last equality we set $m_\chi=100$ GeV. The effective
scattering amplitude is parametrically given by
\begin{equation}
{\cal A}\sim A c_1^p+ v_T c_7^p+\frac{q}{m_N} c_9^p,
\end{equation}
where $v_T\sim 10^{-3}$, $q/m_N\lesssim 0.1$, and $A\sim 100$ for
heavy nuclei. The loop-induced weak contributions thus dominate the
scattering rates of weak scale DM.

\section{Connecting to the physics above the weak scale \label{sec:UV}}

We now describe how to apply and extend our results for the case in
which there is a separation between the mediator scale and the
electroweak scale, i.e., if $\Lambda \gg m_Z$. In this case, the
effective Lagrangian valid above the weak scale is
\begin{equation}
{\lag}_\chi=\sum_{a,d} \frac{C_{a}^{(d)}}{\Lambda^{d-4}} Q_a^{(d)}\,,
\end{equation}
with the operators $Q_a^{(d)}$ manifestly invariant under the full SM
gauge group. We focus on the dimension-six effective interactions
involving DM and quarks currents, analogous to those in
Eqs.~\eqref{eq:dim6:Q1Q2}--\eqref{eq:dim6:Q3Q4}.  For the case of
Dirac-fermion DM in a generic $SU(2)_L$ representation with generators
$\tilde \tau^a$ and hypercharge $Y_\chi$, the basis of dimension-six
operators is \cite{future:BBGZ}
\begin{align}
Q_{1,i}^{(6)} &= (\bar\chi\gamma_\mu \tilde\tau^a\chi)(\bar Q_L^i \gamma^\mu \tau^a Q_L^i)\,,& 
Q_{5,i}^{(6)} &= (\bar\chi\gamma_\mu \gamma_5 \tilde\tau^a\chi)(\bar Q_L^i \gamma^\mu \tau^a Q_L^i)\,, \label{eq:dim6:Q15}\\
Q_{2,i}^{(6)} &= (\bar\chi\gamma_\mu \chi)(\bar Q_L^i \gamma^\mu Q_L^i)\,, & 
Q_{6,i}^{(6)} &= (\bar\chi\gamma_\mu \gamma_5 \chi)(\bar Q_L^i \gamma^\mu Q_L^i)\,,\label{eq:dim6:Q26}\\
Q_{3,i}^{(6)} &= (\bar\chi\gamma_\mu \chi)(\bar u_R^i \gamma^\mu u_R^i)\,, & 
Q_{7,i}^{(6)} &= (\bar\chi\gamma_\mu \gamma_5 \chi)(\bar u_R^i \gamma^\mu u_R^i)\,,\label{eq:dim6:Q37}\\
Q_{4,i}^{(6)} &= (\bar\chi\gamma_\mu \chi)(\bar d_R^i \gamma^\mu d_R^i)\,, & 
Q_{8,i}^{(6)} &= (\bar\chi\gamma_\mu \gamma_5 \chi)(\bar d_R^i \gamma^\mu d_R^i)\,, \label{eq:dim6:Q48}
\end{align}
where the index $i=1,2,3$ labels the generation, and $\tau^a =
\sigma^a/2$, with the Pauli matrices $\sigma^a$.  If $\chi$ is an
electroweak singlet, the operators $Q_{1,i}^{(6)}$ and $Q_{5,i}^{(6)}$
do not exist.  Below the weak scale the above operators,
$Q_{n,i}^{(d)}$, match onto the operators ${\cal Q}_{m,f}^{(d)}$ in
Eqs.~\eqref{eq:dim6:Q1Q2}--\eqref{eq:dim6:Q3Q4}.

For certain patterns of Wilson coefficients, DM couples only to
bottom- and/or charm-quark axial-vector currents.  Such possibilities
are the main focus of this work.  For instance, DM couples (at the
mediator scale $\mu\simeq \Lambda \gg m_Z$) only to the axial
bottom-quark current if the only nonzero Wilson coefficients are
$C_{5,3}^{(6)}$, $C_{6,3}^{(6)}$, and $C_{8,3}^{(6)}$, and such that
they satisfy the relation
\begin{equation}
\label{eq:relations:bottom}
Y_\chi C_{5,3}^{(6)}=4 C_{6,3}^{(6)}=-2C_{8,3}^{(6)}\,.
\end{equation}
We first derive the leading electroweak contribution to DM--nucleon
scattering rates for this case and then discuss the case in which DM
couples only to charm axial currents.

\begin{figure}
	\includegraphics{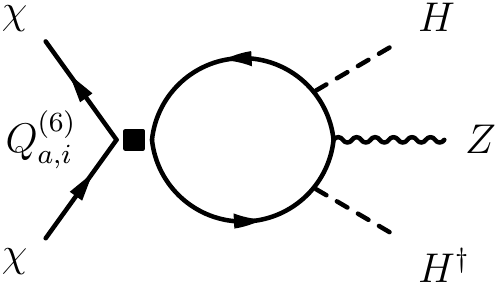}\hspace{5em}
\includegraphics{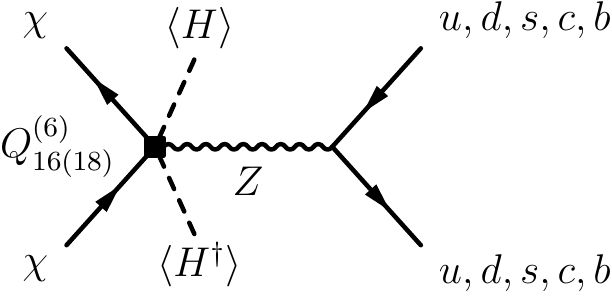}
\caption{Left: ``electroweak fish'' diagram that induces the mixing of
  $Q_{a,i}^{(6)}$ operators into the Higgs operators, $Q_{16}^{(6)}$.
  Right: tree-level diagram that generates the dimension-eight
  operators from the Higgs operators through matching at the weak
  scale.
\label{fig:ew_fish}}
\end{figure}

To this end, we fist assume that the initial conditions at $\mu\simeq
\Lambda$ satisfy Eq.~\eqref{eq:relations:bottom}.  At scales $\muew <
\mu < \Lambda$, the operators $Q_{a,i}^{(6)}$ mix at one-loop via the
SM Yukawa interactions into the Higgs-current operators
\cite{Crivellin:2014qxa, DEramo:2014nmf, future:BBGZ}
\begin{align}
Q_{16}^{(6)} &= (\bar\chi\gamma^\mu \chi)(H^\dagger
i\overset{\leftrightarrow}{D}_\mu H) \,, & Q_{18}^{(6)} &=
(\bar\chi\gamma^\mu \gamma_5 \chi)(H^\dagger
i\overset{\leftrightarrow}{D}_\mu H)\,,
\label{eq:dim6:Q16Q18} 
\end{align} 
and a similar set of operators with the $\tilde \tau^a \otimes \tau^a$
structure (above, $\overset{\leftrightarrow}{D}_\mu\,\equiv\,
D_\mu-\overset{\leftarrow}{D}_\mu$).  This mixing is generated by
``electroweak fish'' diagrams, see Fig.~\ref{fig:ew_fish} (left), and
induces at $\muew\simeq m_Z$
\begin{equation}\label{eq:C16:6}
	C_{16}^{(6)}(m_Z)\sim \frac{y_b^2}{16\pi^2} C^{(6)}_{a,3}(\Lambda) \log\frac{m_Z}{\Lambda}\,, 
\end{equation}
Here, we only show the parametric dependence and suppress ${\cal O}(1)$ coefficients 
from the actual value of the anomalous dimensions.

At energies close to the electroweak scale, at which the Higgs obtains
its vacuum expectation value, the two operators in
Eq.~\eqref{eq:dim6:Q16Q18} result in couplings of DM currents to the
$Z$ boson.  Integrating out the $Z$ boson at tree-level induces DM
couplings to quarks, see Fig.~\ref{fig:ew_fish} (right).  The
Higgs-current operators in Eq.~\eqref{eq:dim6:Q16Q18} therefore match,
at $\muew$, onto four-fermion operators of the five-flavor EFT that
couple DM to quarks with an interaction strength of parametric size
$\sim G_F \, C_{16}^{(6)} (m_Z) \, v^2/\Lambda^2 $.  The factor $v^2$
originates from the two Higgs fields relaxing to their vacuum
expectation values and the factor $G_F$ from integrating out the $Z$
boson.

Since the one-loop RG running from $\Lambda$ to $\muew$,
Eq. \eqref{eq:C16:6}, followed by tree-level matching at $\muew$,
induces interactions proportional to $y_b^2 v^2$, it is convenient to
match such corrections to initial conditions of the dimension-eight
operators in Eqs.~\eqref{eq:dim8:Q34}.\footnote{Here, we have decided
  to ascribe the tree-level $Z$ exchange contribution from the
  matching at $\muew$ to dimension-eight four-fermion
  operators. Alternatively, we could have absorbed also this
  contribution into the Wilson coefficients of dimension-six
  operators. This choice would have the unattractive property of
  having the parametric suppression of $y_b^2 v^2 G_F=m_b^2G_F$ hidden
  in the smallness of some of (the parts of) the Wilson coefficients
  ${\cal C}^{(6)}_{a,f}$, thus making the five-flavor EFT less
  transparent. With our choice, the parametric suppression of
  $m_b^2G_F$ is factored out of the Wilson coefficients and is part of
  the mass-dimension counting of the dimension-eight operator.}  For
the pattern of initial conditions in Eq.~\eqref{eq:relations:bottom},
we then find that the Wilson coefficient of the dimension-eight
operator ${\cal Q}^{(8)}_{4,b}$ at $\muew\simeq m_Z$ is
\begin{equation}
\label{eq:C84b:mZ}
{\cal C}^{(8)}_{4,b} (m_Z) \sim \frac{g_s^2}{16\pi^2}
C_{a,3}^{(6)}(\Lambda) \log \frac{m_Z}{\Lambda}\,,
\end{equation}
and the Wilson coefficient of the dimension-six operator 
${\cal  Q}^{(6)}_{4,b}$ is
\begin{equation}
{\cal C}^{(6)}_{4,b} (m_Z) \sim C_{a,3}^{(6)}(\Lambda) \,.
\end{equation} 
Again, we only show the parametric dependence, including loop factors,
but omit ${\mathcal O}(1)$ factors, e.g., from the actual values of
anomalous dimensions (for details see Ref.~\cite{future:BBGZ}).  In
particular, $C_{a,3}^{(6)}(\Lambda)$ denotes a linear combination of
the Wilson coefficients with $a=5,6,8$.

The subsequent RG evolution from $\muew$ to $\mustr$ proceeds as
described in Section \ref{sec:match},
Eqs.~\eqref{eq:rge:eff}--\eqref{eq:matching:heavyquark}.  The only
difference is that the initial conditions ${\cal C}^{(8)}_{4,b} (m_Z)$
are now nonzero.  For instance, in the result for the non-relativistic
coefficient $c_4^p$ in Eq.~\eqref{eq:cNR4:exp}, there is an additional
contribution from ${\cal C}^{(8)}_{4,b} (m_Z)\propto \log
({m_Z}/{\Lambda})$.  If one neglects the QCD effects, the two
contributions amount to adding up two logarithmically enhanced terms
with exactly the same prefactors.  The net effect is to replace
$\log(m_{q'}/m_Z)$ with $\log(m_{q'}/\Lambda)=\log(m_{q'}/m_Z) +
\log(m_Z/\Lambda)$ in Eqs.~\eqref{eq:cNR1:exp}--\eqref{eq:cNR9:exp}.
This is equivalent to simply calculating the electroweak fish diagram,
with $u,d,s$ fermions attached to the $Z$ line, and keeping only the
$\log(m_b/\Lambda)$-enhanced part.

\bigskip
An analogous analysis applies if the only nonzero Wilson coefficients
satisfy $Y_\chi C_{5,2}^{(6)}=-4 C_{6,2}^{(6)}=2C_{7,2}^{(6)}$, so
that just the $(\bar \chi \gamma_\mu\gamma_5 \chi)(\bar c
\gamma^\mu\gamma_5 c)$ operator is generated (setting the CKM matrix
to unity for simplicity). Similarly, if the only nonzero Wilson
coefficients satisfy $Y_\chi C_{1,3}^{(6)}=4
C_{2,3}^{(6)}=-2C_{4,3}^{(6)}\ne 0$ or $Y_\chi C_{1,2}^{(6)} = -4
C_{2,2}^{(6)}=2C_{3,2}^{(6)}\ne 0$ this means that just the $(\bar
\chi \gamma_\mu \chi)(\bar b \gamma^\mu\gamma_5 b)$ or $(\bar \chi
\gamma_\mu \chi)(\bar c \gamma^\mu\gamma_5 c)$ operators are
generated. Such relations do not necessarily imply fine-tuning, as
they can originate from the quantum number assignments for the
mediators, DM, and quark fields in the UV theory. They do require the
DM hypercharge $Y_\chi$ to be nonzero.\footnote{This may or may not
  lead to potentially dangerous renormalizable couplings of DM to the
  $Z$-boson. An example of the latter is a DM multiplet that is a
  pseudo-Dirac fermion (a Dirac fermion with an additional small
  Majorana mass term), such as an almost pure Higgsino in the MSSM.
  In this case, the lightest mass eigenstate, the Majorana-fermion DM,
  does not couple diagonally to the $Z$ boson at tree level.}  This
conclusion changes, if at $\mu\simeq \Lambda$ we also include
dimension-eight operators of the form $(\bar \chi \gamma_\mu
\chi)(\bar Q_L H \gamma^\mu H Q_L)$ alongside the dimension-six $(\bar
\chi \gamma_\mu \chi)(\bar b_R \gamma^\mu b_R)$ operators. In this
case, it is possible to induce only the $(\bar \chi \gamma_\mu
\chi)(\bar b \gamma^\mu\gamma_5 b)$ or $(\bar \chi \gamma_\mu
\chi)(\bar c \gamma^\mu\gamma_5 c)$ operators even for DM with zero
hypercharge (and thus without a renormalizable interaction to the $Z$
boson).  This, however, requires fine-tuning of dimension-six and
dimension-eight contributions.

Note that the relation in Eq.~\eqref{eq:relations:bottom} also
requires DM to be part of an electroweak multiplet.  For singlet DM
there is no operator $Q_{5,i}^{(6)}$ and so $C_{5,i}^{(6)}$ is
trivially zero. Therefore, for singlet DM a coupling to an
axial-vector bottom-quark current is always accompanied by couplings
to top quarks.  In this case our results get corrected by terms of
order $y_t^2\log(\muew/\Lambda)$ from the RG evolution above the
electroweak scale due to top-Yukawa interactions \cite{future:BBGZ}.

Another phenomenologically interesting case is the one of 
DM coupling only to leptons at $\muew$, i.e., through operators in
Eqs.~\eqref{eq:dim6:Q1Q2}--\eqref{eq:dim6:Q3Q4} with $f=e,\mu,\tau$.
We can readily adapt our results to this case by replacing in
Eqs.~\eqref{eq:op:qq':12}--\eqref{eq:op:q:34} the bottom- and
charm-quarks with leptons.  The new operators are either
color-singlets or conserved QCD currents so that their anomalous
dimensions vanish.  The hadronic functions $c_i^{p(n)}$, controlling
DM scattering on nuclei, are then given by
Eqs.~\eqref{eq:cNR1:exp}--\eqref{eq:cNR9:exp} after substituting
$q'\to {\ell=\tau,\mu}$, and dividing by the number of colors,
$N_c=3$, implicit in these equations.  For $\ell=\mu,\tau$ there are
no other numerically competing contributions from other
dimension-eight operators apart from the ones we discussed here.
However, for the electron case, $\ell=e$, we expect dimension-eight
operators with derivatives, which we have not considered here, to
contribute to the scattering on nuclei at approximately the same
order.

\section{Majorana and scalar dark matter\label{sec:other}}

So far we focused on DM that is a Dirac fermion.  However, the RG
results discussed in this work do not depend on the precise form of
the DM current.  We can, therefore, generalize our results to the case
of Majorana and scalar DM.

\subsection{Majorana dark matter}

Our results apply for Majorana DM with only small modifications.  It
is sufficient to drop from the operator basis the operators ${\cal
  Q}_{1,f}^{(6)}$ and ${\cal Q}_{3,f}^{(6)}$ in
Eqs.~\eqref{eq:dim6:Q1Q2}--\eqref{eq:dim6:Q3Q4} and likewise the
operators ${\cal Q}_{1,q}^{(8)}$ and ${\cal Q}_{3,q}^{(8)}$ in
Eqs.~\eqref{eq:dim8:Q12}--\eqref{eq:dim8:Q34}.  The Lagrangian terms
proportional to the remaining operators, ${\cal Q}_{2,f}^{(6)}$,
${\cal Q}_{4,f}^{(6)}$, ${\cal Q}_{2,f}^{(8)}$, and ${\cal
  Q}_{4,f}^{(8)}$ should be multiplied by a factor of $1/2$ to account
for the additional Wick contractions (see, for instance,
Ref.~\cite{Bishara:2017nnn}).  With these modifications, the
coefficients of the nuclear effective theory are still given by
Eqs.~\eqref{eq:cNR1}--\eqref{eq:cNR9}.

\subsection{Scalar dark matter}
The relevant set of operators for scalar DM is 
\begin{align}
\label{eq:dim6:Q1Q2:light:scalar}
{\cal Q}_{1,f}^{(6)}&=\big(\varphi^* i\overset{\!\!\leftrightarrow}{\partial_\mu} \varphi\big) (\bar f \gamma^\mu f)\,,&
{\cal Q}_{2,f}^{(6)}&=\big(\varphi^* i\overset{\!\!\leftrightarrow}{\partial_\mu} \varphi\big)(\bar f \gamma^\mu \gamma_5 f)\,,
\end{align}
where $\varphi^* \overset{\!\!\leftrightarrow}{\partial_\mu} \varphi
\equiv \varphi^* \partial_\mu \varphi- (\partial_\mu \varphi^*)
\varphi$. These operators are part of the dimension-six effective
Lagrangian for scalar DM, cf., Ref.~\cite{Bishara:2016hek},
\begin{equation}\label{eq:lightDM:Ln:scalar}
{\lag}_\varphi =
\sum_a \frac{\C_{a}^{(6)}}{\Lambda^{2}} {\cal Q}_a^{(6)} \,,
\end{equation}
with $\C_{a}^{(6)}$ the dimensionless Wilson coefficients. Note that
we adopt the same notation for operators and Wilson coefficients in
the case of scalar DM as we did for fermionic DM.  No confusion should
arise as this abuse of notation is restricted to this subsection.

Apart from having a different DM current, nothing changes in our
calculations. Therefore, after defining the dimension-eight effective
Lagrangian in the three-flavor theory as 
\begin{align}
\label{eq:dim8:Q1Q2:light:scalar}
{\lag}_\varphi^{(8)} = - \frac{\sqrt 2 G_F}{\Lambda^2}\sum_{q=u,d,s}
\Big( {\cal C}_{1,q}^{(8)} \frac{m_q^2}{g_s^2} \big(\varphi^*
i\overset{\!\!\leftrightarrow}{\partial_\mu} \varphi\big) (\bar q
\gamma^\mu q) 
 + {\cal C}_{2,f}^{(8)} \frac{m_q^2}{g_s^2} \big(\varphi^*
 i\overset{\!\!\leftrightarrow}{\partial_\mu} \varphi\big)(\bar q
 \gamma^\mu \gamma_5 q) \Big)\,, 
\end{align}
the additional contributions to the nuclear coefficients are given,
for complex scalar DM, by (cf. Ref.~\cite{Bishara:2017nnn})
\begin{align}
c_1^N&= - \frac{1}{2\pi \alpha_s} \frac{\sqrt{2}G_F}{ \Lambda^2} m_\varphi \sum_{q=u,d,s} 
m_q^2\, 
\C_{1q}^{(8)}\,
F_1^{q/N}\,, 
\\
c_7^N&=\phantom{+}~\, \frac{1}{\pi\alpha_s}\frac{\sqrt{2}G_F}{\Lambda^2} m_\varphi \sum_{q=u,d,s} 
m_q^2\, 
\C_{2q}^{(8)}\,
F_A^{q/N}\,. 
\end{align}
For real scalar DM, the operators in
Eq.~\eqref{eq:dim6:Q1Q2:light:scalar} vanish.  For completeness, we
display also the dimension-eight contributions to the nuclear
coefficients, expanded to leading order in the strong coupling
constant,
\begin{align}
\label{eq:cNR1:exp:scalar}
c_{1}^N \simeq&\phantom{+}~\,~\frac{6m_\varphi}{\pi^2} \frac{\sqrt{2} G_F}{\Lambda^2} \sum_{q'=c,b} 
m_{q'}^2 \, 
\log\frac{\mu_{q'}}{m_Z}\,
\C_{2,q'}^{(6)}(m_Z)\,
\sum_{q=u,d,s} {\cal D}_{3,q'q}F_{1}^{q/N} +{\mathcal O}(\alpha_s) \,, 
\\
\label{eq:cNR7:exp:scalar}
c_{7}^N \simeq& - \frac{12m_\varphi}{\pi^2}\frac{\sqrt{2} G_F}{\Lambda^2} \sum_{q'=c,b} 
m_{q'}^2  \, 
\log\frac{\mu_{q'}}{m_Z}\,
\C_{2,q'}^{(6)}(m_Z) \,
\sum_{q=u,d,s} {\cal D}_{2,qq'}  F_{A}^{q/N}  +{\mathcal O}(\alpha_s) \,.
\end{align}

\section{Conclusions\label{sec:discussion}}

If DM couples only to bottom- or charm-quark axial-vector currents,
the dominant contribution to DM scattering on nuclei is either due to
one-loop electroweak corrections or due to the intrinsic bottom and
charm axial charges of the nucleons. Below the weak scale the
electroweak contributions are captured by double insertions of both
the DM effective Lagrangian and the SM weak effective
Lagrangian. These convert the heavy-quark currents to the currents
with $u,d$, and $s$ quarks that have nonzero nuclear matrix
elements. In this paper we calculated the nonrelativistic couplings of
DM to neutrons and protons that result from such electroweak
corrections, including the resummation of the leading-logarithmic QCD
corrections.  The latter are numerically important, as they lead to
${\mathcal O}(1)$ changes in the scattering rates on nuclei.  Our
results can be readily included in the general framework of EFT for DM
direct detection, and will be implemented in a future version of the
public code \texttt{DirectDM}~\cite{Bishara:2017nnn}.

\bigskip
{\bf Acknowledgements.} We thank Francesco D'Eramo for useful
discussions, and especially Fady Bishara for checking several
equations.  JZ acknowledges support in part by the DOE grant
DE-SC0011784. The research of BG was supported in part by the DOE Grant No. DE-SC0009919.

\appendix
\section{Unphysical operators\label{App:unphysical}}
We extract the anomalous dimensions by renormalizing off-shell Green's
functions in $d=4-2\epsilon$ dimensions.  In some intermediate stages
of the computation it is thus necessary to introduce some unphysical
operators.

\subsection{Evanescent operators}
The one-loop mixing among the ``physical'' operators is not affected
by the definition of {\it evanescent} operators, i.e., operators that
are required to project one-loop Green's functions in $d=4-2\epsilon$
dimensions but vanish in $d=4$.  Indeed, our one-loop results could
also have been obtained by performing the Dirac algebra in $d=4$
instead off in non-integer dimensions.  Since {\it i)} this no longer
possible at next-to-leading order computations and {\it ii)} we use
dimensional regularization to extract the poles of loop integrals, we
find it convenient to also perform the Dirac algebra in non-integer
dimensions.  To project the $d=4-2\epsilon$ amplitudes we thus need to
also include some evanescent operators in the basis.  For completeness
and future reference, we list below the ones entering the one-loop
computations.
\begin{align} 
\label{eq:evan:1:LO}
\begin{split}
{\cal E}_1^{qq'} & = (\bar q \gamma^\mu \gamma^\nu \gamma^\rho q) \, (\bar q'
\gamma_\mu \gamma_\nu \gamma_\rho q') - 10 {\cal O}_1^{qq'} - 6 {\cal O}_2^{qq'} \,, \\[0.5em]
{\cal E}_2^{qq'} & =
(\bar q \gamma^\mu \gamma^\nu \gamma^\rho \gamma_5 q) \, (\bar q'
\gamma_\mu \gamma_\nu \gamma_\rho \gamma_5 q') - 6 {\cal O}_1^{qq'} - 10 {\cal O}_2^{qq'} \,, \\[0.5em] 
{\cal E}_3^{qq'} & = (\bar q \gamma^\mu \gamma^\nu \gamma^\rho \gamma_5 q) \, (\bar q'
\gamma_\mu \gamma_\nu \gamma_\rho q') - 10 {\cal O}_3^{qq'} - 6 {\cal O}_3^{q'q} \,, \\[0.5em] 
{\cal E}_4^{qq'} & = (\bar q \gamma^\mu \gamma^\nu \gamma^\rho \, T^a q) \, (\bar q'
\gamma_\mu \gamma_\nu \gamma_\rho \, T^a q') - 10 {\cal O}_4^{qq'} - 6 {\cal O}_5^{q'q} \,, \\[0.5em] 
{\cal E}_5^{qq'} & = (\bar q \gamma^\mu \gamma^\nu \gamma^\rho \gamma_5 \, T^a q) \, (\bar q'
\gamma_\mu \gamma_\nu \gamma_\rho \gamma_5 \, T^a q') - 6 {\cal O}_4^{qq'} - 10 {\cal O}_5^{q'q} \,, \\[0.5em] 
{\cal E}_6^{qq'} & = (\bar q \gamma^\mu \gamma^\nu \gamma^\rho \gamma_5 \, T^a q) \, (\bar q'
\gamma_\mu \gamma_\nu \gamma_\rho \, T^a q') - 10 {\cal O}_6^{qq'} - 6 {\cal O}_6^{q'q} \,,
\end{split}\\[0.5em]
\intertext{and}
\label{eq:evan:2:LO}
\begin{split}
{\cal E}_1^{q} & = (\bar q \gamma^\mu \gamma^\nu \gamma^\rho q) \, (\bar q
\gamma_\mu \gamma_\nu \gamma_\rho q) - 10 {\cal O}_1^{q} - 6 {\cal O}_2^{q} \,, \\[0.5em]
{\cal E}_2^{q} & = (\bar q \gamma^\mu \gamma^\nu \gamma^\rho \gamma_5 q) \, (\bar q
\gamma_\mu \gamma_\nu \gamma_\rho \gamma_5 q) - 6 {\cal O}_1^{q} - 10 {\cal O}_2^{q} \,, \\[0.5em] 
{\cal E}_3^{q} & = (\bar q \gamma^\mu \gamma^\nu \gamma^\rho \gamma_5 q) \, (\bar q
\gamma_\mu \gamma_\nu \gamma_\rho q) - 16 {\cal O}_3^{q} \,, \\[0.5em] 
{\cal E}_4^{q} & = 
(\bar q \gamma^\mu \gamma^\nu \gamma^\rho \, T^a q) \, (\bar q
\gamma_\mu \gamma_\nu \gamma_\rho \, T^a q) - 3 \bigg( 1 - \frac{1}{N_c} \bigg) \big( {\cal
  O}_1^{q} + {\cal O}_2^{q} \big) - 4 {\cal O}_4^{q} \,, \\[0.5em] 
{\cal E}_5^{q} & = (\bar q \gamma^\mu \gamma^\nu \gamma^\rho \gamma_5 \, T^a q) \, (\bar q
\gamma_\mu \gamma_\nu \gamma_\rho \gamma_5 \, T^a q) - 5 \bigg( 1 - \frac{1}{N_c} \bigg) \big( {\cal
  O}_1^{q} + {\cal O}_2^{q} \big) + 4 {\cal O}_4^{q} \,, \\[0.5em] 
{\cal E}_6^{q} & = (\bar q \gamma^\mu \gamma^\nu \gamma^\rho \gamma_5 \, T^a q) \, (\bar q
\gamma_\mu \gamma_\nu \gamma_\rho \, T^a q) - 8 \bigg( 1 - \frac{1}{N_c} \bigg) {\cal O}_3^{q} \,, \\[0.5em] 
{\cal E}_7^{q} & = 
(\bar q \gamma^\mu \gamma_5 \, T^a q) \, (\bar q
\gamma_\mu \gamma_5 \, T^a q) - \frac{1}{2}\bigg( 1 - \frac{1}{N_c} \bigg) \big( {\cal
  O}_1^{q} +{\cal O}_2^{q} \big) + {\cal O}_4^{q} \,, \\[0.5em] 
{\cal E}_8^{q} & = (\bar q \gamma^\mu \gamma_5 \, T^a q) \, (\bar q
\gamma_\mu \, T^a q) - \frac{1}{2} \bigg( 1 - \frac{1}{N_c} \bigg) {\cal O}_3^{q} \,.
\end{split}
\end{align}
Here, $N_c=3$ denotes the number of colors.
The operators ${\cal E}^q_7$ and ${\cal E}^q_8$ are Fierz-evanenscent operators,
i.e., they vanish due to Fierz identities and not due to special $d=4$ relations
of the Dirac algebra.

\subsection{E.o.m.-vanishing operators}

In our conventions the equation of motion (e.o.m.) for the gluon 
field reads
\begin{equation}
\begin{split}
D^\nu  G^{a}_{\nu\mu} \equiv (\partial^\nu \delta^{ab} - g_s
f^{abc} G^{\nu,c}) G^{b}_{\nu\mu} = - g_s \sum_q
\bar q T^a \gamma_\mu q \,,
\end{split}
\end{equation}
up to gauge-fixing and ghost terms. The sum is over all active
quark fields. 
Hence the following operators vanish via the e.o.m.
\begin{equation} \label{eq:eom}
\begin{split}
{\cal N}_{1,\text{e.o.m.}}^{q} & = \frac{1}{g_s} (\bar q \gamma^\mu \, T^a
q) D^\nu  G^{a}_{\nu\mu} + {\cal O}_{4,q}^{(6)} + \sum_{q'\neq q} {\cal O}_{4,qq'}^{(6)} \,, \\
{\cal N}_{2,\text{e.o.m.}}^{q} & = \frac{1}{g_s} (\bar q \gamma^\mu \gamma_5 \, T^a
q) D^\nu  G^{a}_{\nu\mu} + \frac{1}{2}\bigg( 1 - \frac{1}{N_c} \bigg) {\cal O}_3^{q} + \sum_{q'\neq q} {\cal O}_{6,qq'}^{(6)} \,.
\end{split}
\end{equation}
The four-fermion pieces of these e.o.m.-vanishing operators contribute
to the same amplitudes as the physical four-fermion
operators. Therefore, the mixing of physical operators into the
e.o.m.-vanishing operators (computed from QCD penguin diagrams,
Fig.~\ref{fig:4fermionrunning}) affects the anomalous dimensions of
four-fermion operators.

\bibliography{paper}

\begin{thebibliography}{38}
\expandafter\ifx\csname natexlab\endcsname\relax\def\natexlab#1{#1}\fi
\expandafter\ifx\csname bibnamefont\endcsname\relax
  \def\bibnamefont#1{#1}\fi
\expandafter\ifx\csname bibfnamefont\endcsname\relax
  \def\bibfnamefont#1{#1}\fi
\expandafter\ifx\csname citenamefont\endcsname\relax
  \def\citenamefont#1{#1}\fi
\expandafter\ifx\csname url\endcsname\relax
  \def\url#1{\texttt{#1}}\fi
\expandafter\ifx\csname urlprefix\endcsname\relax\def\urlprefix{URL }\fi
\providecommand{\bibinfo}[2]{#2}
\providecommand{\eprint}[2][]{\url{#2}}

\bibitem[{\citenamefont{Bagnasco et~al.}(1994)\citenamefont{Bagnasco, Dine, and
  Thomas}}]{Bagnasco:1993st}
\bibinfo{author}{\bibfnamefont{J.}~\bibnamefont{Bagnasco}},
  \bibinfo{author}{\bibfnamefont{M.}~\bibnamefont{Dine}}, \bibnamefont{and}
  \bibinfo{author}{\bibfnamefont{S.~D.} \bibnamefont{Thomas}},
  \bibinfo{journal}{Phys. Lett.} \textbf{\bibinfo{volume}{B320}},
  \bibinfo{pages}{99} (\bibinfo{year}{1994}), \eprint{hep-ph/9310290}.

\bibitem[{\citenamefont{Kurylov and Kamionkowski}(2004)}]{Kurylov:2003ra}
\bibinfo{author}{\bibfnamefont{A.}~\bibnamefont{Kurylov}} \bibnamefont{and}
  \bibinfo{author}{\bibfnamefont{M.}~\bibnamefont{Kamionkowski}},
  \bibinfo{journal}{Phys. Rev.} \textbf{\bibinfo{volume}{D69}},
  \bibinfo{pages}{063503} (\bibinfo{year}{2004}), \eprint{hep-ph/0307185}.

\bibitem[{\citenamefont{Kopp et~al.}(2010)\citenamefont{Kopp, Schwetz, and
  Zupan}}]{Kopp:2009qt}
\bibinfo{author}{\bibfnamefont{J.}~\bibnamefont{Kopp}},
  \bibinfo{author}{\bibfnamefont{T.}~\bibnamefont{Schwetz}}, \bibnamefont{and}
  \bibinfo{author}{\bibfnamefont{J.}~\bibnamefont{Zupan}},
  \bibinfo{journal}{JCAP} \textbf{\bibinfo{volume}{1002}}, \bibinfo{pages}{014}
  (\bibinfo{year}{2010}), \eprint{0912.4264}.

\bibitem[{\citenamefont{Goodman et~al.}(2011)\citenamefont{Goodman, Ibe,
  Rajaraman, Shepherd, Tait, and Yu}}]{Goodman:2010qn}
\bibinfo{author}{\bibfnamefont{J.}~\bibnamefont{Goodman}},
  \bibinfo{author}{\bibfnamefont{M.}~\bibnamefont{Ibe}},
  \bibinfo{author}{\bibfnamefont{A.}~\bibnamefont{Rajaraman}},
  \bibinfo{author}{\bibfnamefont{W.}~\bibnamefont{Shepherd}},
  \bibinfo{author}{\bibfnamefont{T.~M.~P.} \bibnamefont{Tait}},
  \bibnamefont{and} \bibinfo{author}{\bibfnamefont{H.-B.} \bibnamefont{Yu}},
  \bibinfo{journal}{Nucl. Phys.} \textbf{\bibinfo{volume}{B844}},
  \bibinfo{pages}{55} (\bibinfo{year}{2011}), \eprint{1009.0008}.

\bibitem[{\citenamefont{Goodman et~al.}(2010)\citenamefont{Goodman, Ibe,
  Rajaraman, Shepherd, Tait et~al.}}]{Goodman:2010ku}
\bibinfo{author}{\bibfnamefont{J.}~\bibnamefont{Goodman}},
  \bibinfo{author}{\bibfnamefont{M.}~\bibnamefont{Ibe}},
  \bibinfo{author}{\bibfnamefont{A.}~\bibnamefont{Rajaraman}},
  \bibinfo{author}{\bibfnamefont{W.}~\bibnamefont{Shepherd}},
  \bibinfo{author}{\bibfnamefont{T.~M.} \bibnamefont{Tait}},
  \bibnamefont{et~al.}, \bibinfo{journal}{Phys.Rev.}
  \textbf{\bibinfo{volume}{D82}}, \bibinfo{pages}{116010}
  (\bibinfo{year}{2010}), \eprint{1008.1783}.

\bibitem[{\citenamefont{Bai et~al.}(2010)\citenamefont{Bai, Fox, and
  Harnik}}]{Bai:2010hh}
\bibinfo{author}{\bibfnamefont{Y.}~\bibnamefont{Bai}},
  \bibinfo{author}{\bibfnamefont{P.~J.} \bibnamefont{Fox}}, \bibnamefont{and}
  \bibinfo{author}{\bibfnamefont{R.}~\bibnamefont{Harnik}},
  \bibinfo{journal}{JHEP} \textbf{\bibinfo{volume}{12}}, \bibinfo{pages}{048}
  (\bibinfo{year}{2010}), \eprint{1005.3797}.

\bibitem[{\citenamefont{Hill and Solon}(2012)}]{Hill:2011be}
\bibinfo{author}{\bibfnamefont{R.~J.} \bibnamefont{Hill}} \bibnamefont{and}
  \bibinfo{author}{\bibfnamefont{M.~P.} \bibnamefont{Solon}},
  \bibinfo{journal}{Phys.Lett.} \textbf{\bibinfo{volume}{B707}},
  \bibinfo{pages}{539} (\bibinfo{year}{2012}), \eprint{1111.0016}.

\bibitem[{\citenamefont{Cirelli et~al.}(2013)\citenamefont{Cirelli, Del~Nobile,
  and Panci}}]{DelNobile:2013sia}
\bibinfo{author}{\bibfnamefont{M.}~\bibnamefont{Cirelli}},
  \bibinfo{author}{\bibfnamefont{E.}~\bibnamefont{Del~Nobile}},
  \bibnamefont{and} \bibinfo{author}{\bibfnamefont{P.}~\bibnamefont{Panci}},
  \bibinfo{journal}{JCAP} \textbf{\bibinfo{volume}{1310}}, \bibinfo{pages}{019}
  (\bibinfo{year}{2013}), \eprint{1307.5955}.

\bibitem[{\citenamefont{Hill and Solon}(2015)}]{Hill:2014yxa}
\bibinfo{author}{\bibfnamefont{R.~J.} \bibnamefont{Hill}} \bibnamefont{and}
  \bibinfo{author}{\bibfnamefont{M.~P.} \bibnamefont{Solon}},
  \bibinfo{journal}{Phys.Rev.} \textbf{\bibinfo{volume}{D91}},
  \bibinfo{pages}{043505} (\bibinfo{year}{2015}), \eprint{1409.8290}.

\bibitem[{\citenamefont{Crivellin et~al.}(2014)\citenamefont{Crivellin,
  D'Eramo, and Procura}}]{Crivellin:2014qxa}
\bibinfo{author}{\bibfnamefont{A.}~\bibnamefont{Crivellin}},
  \bibinfo{author}{\bibfnamefont{F.}~\bibnamefont{D'Eramo}}, \bibnamefont{and}
  \bibinfo{author}{\bibfnamefont{M.}~\bibnamefont{Procura}},
  \bibinfo{journal}{Phys. Rev. Lett.} \textbf{\bibinfo{volume}{112}},
  \bibinfo{pages}{191304} (\bibinfo{year}{2014}), \eprint{1402.1173}.

\bibitem[{\citenamefont{D'Eramo and Procura}(2015)}]{DEramo:2014nmf}
\bibinfo{author}{\bibfnamefont{F.}~\bibnamefont{D'Eramo}} \bibnamefont{and}
  \bibinfo{author}{\bibfnamefont{M.}~\bibnamefont{Procura}},
  \bibinfo{journal}{JHEP} \textbf{\bibinfo{volume}{04}}, \bibinfo{pages}{054}
  (\bibinfo{year}{2015}), \eprint{1411.3342}.

\bibitem[{\citenamefont{Hill and Solon}(2014)}]{Hill:2013hoa}
\bibinfo{author}{\bibfnamefont{R.~J.} \bibnamefont{Hill}} \bibnamefont{and}
  \bibinfo{author}{\bibfnamefont{M.~P.} \bibnamefont{Solon}},
  \bibinfo{journal}{Phys. Rev. Lett.} \textbf{\bibinfo{volume}{112}},
  \bibinfo{pages}{211602} (\bibinfo{year}{2014}), \eprint{1309.4092}.

\bibitem[{\citenamefont{Hoferichter et~al.}(2015)\citenamefont{Hoferichter,
  Klos, and Schwenk}}]{Hoferichter:2015ipa}
\bibinfo{author}{\bibfnamefont{M.}~\bibnamefont{Hoferichter}},
  \bibinfo{author}{\bibfnamefont{P.}~\bibnamefont{Klos}}, \bibnamefont{and}
  \bibinfo{author}{\bibfnamefont{A.}~\bibnamefont{Schwenk}},
  \bibinfo{journal}{Phys. Lett.} \textbf{\bibinfo{volume}{B746}},
  \bibinfo{pages}{410} (\bibinfo{year}{2015}), \eprint{1503.04811}.

\bibitem[{\citenamefont{Bishara
  et~al.}(2017{\natexlab{a}})\citenamefont{Bishara, Brod, Grinstein, and
  Zupan}}]{Bishara:2016hek}
\bibinfo{author}{\bibfnamefont{F.}~\bibnamefont{Bishara}},
  \bibinfo{author}{\bibfnamefont{J.}~\bibnamefont{Brod}},
  \bibinfo{author}{\bibfnamefont{B.}~\bibnamefont{Grinstein}},
  \bibnamefont{and} \bibinfo{author}{\bibfnamefont{J.}~\bibnamefont{Zupan}},
  \bibinfo{journal}{JCAP} \textbf{\bibinfo{volume}{1702}}, \bibinfo{pages}{009}
  (\bibinfo{year}{2017}{\natexlab{a}}), \eprint{1611.00368}.

\bibitem[{\citenamefont{D'Eramo et~al.}(2016)\citenamefont{D'Eramo, Kavanagh,
  and Panci}}]{DEramo:2016gos}
\bibinfo{author}{\bibfnamefont{F.}~\bibnamefont{D'Eramo}},
  \bibinfo{author}{\bibfnamefont{B.~J.} \bibnamefont{Kavanagh}},
  \bibnamefont{and} \bibinfo{author}{\bibfnamefont{P.}~\bibnamefont{Panci}},
  \bibinfo{journal}{JHEP} \textbf{\bibinfo{volume}{08}}, \bibinfo{pages}{111}
  (\bibinfo{year}{2016}), \eprint{1605.04917}.

\bibitem[{\citenamefont{Bishara
  et~al.}(2017{\natexlab{b}})\citenamefont{Bishara, Brod, Grinstein, and
  Zupan}}]{Bishara:2017pfq}
\bibinfo{author}{\bibfnamefont{F.}~\bibnamefont{Bishara}},
  \bibinfo{author}{\bibfnamefont{J.}~\bibnamefont{Brod}},
  \bibinfo{author}{\bibfnamefont{B.}~\bibnamefont{Grinstein}},
  \bibnamefont{and} \bibinfo{author}{\bibfnamefont{J.}~\bibnamefont{Zupan}},
  \bibinfo{journal}{JHEP} \textbf{\bibinfo{volume}{11}}, \bibinfo{pages}{059}
  (\bibinfo{year}{2017}{\natexlab{b}}), \eprint{1707.06998}.

\bibitem[{\citenamefont{D'Eramo et~al.}(2017)\citenamefont{D'Eramo, Kavanagh,
  and Panci}}]{DEramo:2017zqw}
\bibinfo{author}{\bibfnamefont{F.}~\bibnamefont{D'Eramo}},
  \bibinfo{author}{\bibfnamefont{B.~J.} \bibnamefont{Kavanagh}},
  \bibnamefont{and} \bibinfo{author}{\bibfnamefont{P.}~\bibnamefont{Panci}},
  \bibinfo{journal}{Phys. Lett.} \textbf{\bibinfo{volume}{B771}},
  \bibinfo{pages}{339} (\bibinfo{year}{2017}), \eprint{1702.00016}.

\bibitem[{\citenamefont{Bishara
  et~al.}(2017{\natexlab{c}})\citenamefont{Bishara, Brod, Grinstein, and
  Zupan}}]{Bishara:2017nnn}
\bibinfo{author}{\bibfnamefont{F.}~\bibnamefont{Bishara}},
  \bibinfo{author}{\bibfnamefont{J.}~\bibnamefont{Brod}},
  \bibinfo{author}{\bibfnamefont{B.}~\bibnamefont{Grinstein}},
  \bibnamefont{and} \bibinfo{author}{\bibfnamefont{J.}~\bibnamefont{Zupan}}
  (\bibinfo{year}{2017}{\natexlab{c}}), \eprint{1708.02678}.

\bibitem[{\citenamefont{Brod et~al.}(2017)\citenamefont{Brod,
  Gootjes-Dreesbach, Tammaro, and Zupan}}]{Brod:2017bsw}
\bibinfo{author}{\bibfnamefont{J.}~\bibnamefont{Brod}},
  \bibinfo{author}{\bibfnamefont{A.}~\bibnamefont{Gootjes-Dreesbach}},
  \bibinfo{author}{\bibfnamefont{M.}~\bibnamefont{Tammaro}}, \bibnamefont{and}
  \bibinfo{author}{\bibfnamefont{J.}~\bibnamefont{Zupan}}
  (\bibinfo{year}{2017}), \eprint{1710.10218}.

\bibitem[{\citenamefont{Fitzpatrick et~al.}(2013)\citenamefont{Fitzpatrick,
  Haxton, Katz, Lubbers, and Xu}}]{Fitzpatrick:2012ix}
\bibinfo{author}{\bibfnamefont{A.~L.} \bibnamefont{Fitzpatrick}},
  \bibinfo{author}{\bibfnamefont{W.}~\bibnamefont{Haxton}},
  \bibinfo{author}{\bibfnamefont{E.}~\bibnamefont{Katz}},
  \bibinfo{author}{\bibfnamefont{N.}~\bibnamefont{Lubbers}}, \bibnamefont{and}
  \bibinfo{author}{\bibfnamefont{Y.}~\bibnamefont{Xu}}, \bibinfo{journal}{JCAP}
  \textbf{\bibinfo{volume}{1302}}, \bibinfo{pages}{004} (\bibinfo{year}{2013}),
  \eprint{1203.3542}.

\bibitem[{\citenamefont{Fitzpatrick et~al.}(2012)\citenamefont{Fitzpatrick,
  Haxton, Katz, Lubbers, and Xu}}]{Fitzpatrick:2012ib}
\bibinfo{author}{\bibfnamefont{A.~L.} \bibnamefont{Fitzpatrick}},
  \bibinfo{author}{\bibfnamefont{W.}~\bibnamefont{Haxton}},
  \bibinfo{author}{\bibfnamefont{E.}~\bibnamefont{Katz}},
  \bibinfo{author}{\bibfnamefont{N.}~\bibnamefont{Lubbers}}, \bibnamefont{and}
  \bibinfo{author}{\bibfnamefont{Y.}~\bibnamefont{Xu}} (\bibinfo{year}{2012}),
  \eprint{1211.2818}.

\bibitem[{\citenamefont{Anand et~al.}(2014)\citenamefont{Anand, Fitzpatrick,
  and Haxton}}]{Anand:2013yka}
\bibinfo{author}{\bibfnamefont{N.}~\bibnamefont{Anand}},
  \bibinfo{author}{\bibfnamefont{A.~L.} \bibnamefont{Fitzpatrick}},
  \bibnamefont{and} \bibinfo{author}{\bibfnamefont{W.~C.}
  \bibnamefont{Haxton}}, \bibinfo{journal}{Phys. Rev.}
  \textbf{\bibinfo{volume}{C89}}, \bibinfo{pages}{065501}
  (\bibinfo{year}{2014}), \eprint{1308.6288}.

\bibitem[{\citenamefont{Hoferichter et~al.}(2016)\citenamefont{Hoferichter,
  Klos, Men\'{e}ndez, and Schwenk}}]{Hoferichter:2016nvd}
\bibinfo{author}{\bibfnamefont{M.}~\bibnamefont{Hoferichter}},
  \bibinfo{author}{\bibfnamefont{P.}~\bibnamefont{Klos}},
  \bibinfo{author}{\bibfnamefont{J.}~\bibnamefont{Men\'{e}ndez}},
  \bibnamefont{and} \bibinfo{author}{\bibfnamefont{A.}~\bibnamefont{Schwenk}},
  \bibinfo{journal}{Phys. Rev.} \textbf{\bibinfo{volume}{D94}},
  \bibinfo{pages}{063505} (\bibinfo{year}{2016}), \eprint{1605.08043}.

\bibitem[{\citenamefont{Bali et~al.}(2012)}]{QCDSF:2011aa}
\bibinfo{author}{\bibfnamefont{G.~S.} \bibnamefont{Bali}} \bibnamefont{et~al.}
  (\bibinfo{collaboration}{QCDSF}), \bibinfo{journal}{Phys. Rev. Lett.}
  \textbf{\bibinfo{volume}{108}}, \bibinfo{pages}{222001}
  (\bibinfo{year}{2012}), \eprint{1112.3354}.

\bibitem[{\citenamefont{Engelhardt}(2012)}]{Engelhardt:2012gd}
\bibinfo{author}{\bibfnamefont{M.}~\bibnamefont{Engelhardt}},
  \bibinfo{journal}{Phys. Rev.} \textbf{\bibinfo{volume}{D86}},
  \bibinfo{pages}{114510} (\bibinfo{year}{2012}), \eprint{1210.0025}.

\bibitem[{\citenamefont{Bhattacharya et~al.}(2014)\citenamefont{Bhattacharya,
  Gupta, and Yoon}}]{Bhattacharya:2015gma}
\bibinfo{author}{\bibfnamefont{T.}~\bibnamefont{Bhattacharya}},
  \bibinfo{author}{\bibfnamefont{R.}~\bibnamefont{Gupta}}, \bibnamefont{and}
  \bibinfo{author}{\bibfnamefont{B.}~\bibnamefont{Yoon}},
  \bibinfo{journal}{PoS} \textbf{\bibinfo{volume}{LATTICE2014}},
  \bibinfo{pages}{141} (\bibinfo{year}{2014}), \eprint{1503.05975}.

\bibitem[{\citenamefont{Alexandrou et~al.}(2017)\citenamefont{Alexandrou,
  Constantinou, Hadjiyiannakou, Jansen, Kallidonis, Koutsou, and Vaquero
  Aviles-Casco}}]{Alexandrou:2017hac}
\bibinfo{author}{\bibfnamefont{C.}~\bibnamefont{Alexandrou}},
  \bibinfo{author}{\bibfnamefont{M.}~\bibnamefont{Constantinou}},
  \bibinfo{author}{\bibfnamefont{K.}~\bibnamefont{Hadjiyiannakou}},
  \bibinfo{author}{\bibfnamefont{K.}~\bibnamefont{Jansen}},
  \bibinfo{author}{\bibfnamefont{C.}~\bibnamefont{Kallidonis}},
  \bibinfo{author}{\bibfnamefont{G.}~\bibnamefont{Koutsou}}, \bibnamefont{and}
  \bibinfo{author}{\bibfnamefont{A.}~\bibnamefont{Vaquero Aviles-Casco}}
  (\bibinfo{year}{2017}), \eprint{1705.03399}.

\bibitem[{\citenamefont{Polyakov et~al.}(1999)\citenamefont{Polyakov, Schafer,
  and Teryaev}}]{Polyakov:1998rb}
\bibinfo{author}{\bibfnamefont{M.~V.} \bibnamefont{Polyakov}},
  \bibinfo{author}{\bibfnamefont{A.}~\bibnamefont{Schafer}}, \bibnamefont{and}
  \bibinfo{author}{\bibfnamefont{O.~V.} \bibnamefont{Teryaev}},
  \bibinfo{journal}{Phys. Rev.} \textbf{\bibinfo{volume}{D60}},
  \bibinfo{pages}{051502} (\bibinfo{year}{1999}), \eprint{hep-ph/9812393}.

\bibitem[{\citenamefont{Witten}(1977)}]{Witten:1976kx}
\bibinfo{author}{\bibfnamefont{E.}~\bibnamefont{Witten}},
  \bibinfo{journal}{Nucl. Phys.} \textbf{\bibinfo{volume}{B122}},
  \bibinfo{pages}{109} (\bibinfo{year}{1977}).

\bibitem[{\citenamefont{Gilman and Wise}(1983)}]{Gilman:1982ap}
\bibinfo{author}{\bibfnamefont{F.~J.} \bibnamefont{Gilman}} \bibnamefont{and}
  \bibinfo{author}{\bibfnamefont{M.~B.} \bibnamefont{Wise}},
  \bibinfo{journal}{Phys. Rev.} \textbf{\bibinfo{volume}{D27}},
  \bibinfo{pages}{1128} (\bibinfo{year}{1983}).

\bibitem[{\citenamefont{Flynn}(1990)}]{Flynn:1989cf}
\bibinfo{author}{\bibfnamefont{J.~M.} \bibnamefont{Flynn}},
  \bibinfo{journal}{Mod. Phys. Lett.} \textbf{\bibinfo{volume}{A5}},
  \bibinfo{pages}{877} (\bibinfo{year}{1990}).

\bibitem[{\citenamefont{Datta et~al.}(1990)\citenamefont{Datta, Frohlich, and
  Paschos}}]{Datta:1989xp}
\bibinfo{author}{\bibfnamefont{A.}~\bibnamefont{Datta}},
  \bibinfo{author}{\bibfnamefont{J.}~\bibnamefont{Frohlich}}, \bibnamefont{and}
  \bibinfo{author}{\bibfnamefont{E.~A.} \bibnamefont{Paschos}},
  \bibinfo{journal}{Z. Phys.} \textbf{\bibinfo{volume}{C46}},
  \bibinfo{pages}{63} (\bibinfo{year}{1990}).

\bibitem[{\citenamefont{Herrlich and Nierste}(1995)}]{Herrlich:1994kh}
\bibinfo{author}{\bibfnamefont{S.}~\bibnamefont{Herrlich}} \bibnamefont{and}
  \bibinfo{author}{\bibfnamefont{U.}~\bibnamefont{Nierste}},
  \bibinfo{journal}{Nucl. Phys.} \textbf{\bibinfo{volume}{B455}},
  \bibinfo{pages}{39} (\bibinfo{year}{1995}), \eprint{hep-ph/9412375}.

\bibitem[{\citenamefont{Herrlich and Nierste}(1996)}]{Herrlich:1996vf}
\bibinfo{author}{\bibfnamefont{S.}~\bibnamefont{Herrlich}} \bibnamefont{and}
  \bibinfo{author}{\bibfnamefont{U.}~\bibnamefont{Nierste}},
  \bibinfo{journal}{Nucl. Phys.} \textbf{\bibinfo{volume}{B476}},
  \bibinfo{pages}{27} (\bibinfo{year}{1996}), \eprint{hep-ph/9604330}.

\bibitem[{\citenamefont{Brod and Zupan}(2014)}]{Brod:2013sga}
\bibinfo{author}{\bibfnamefont{J.}~\bibnamefont{Brod}} \bibnamefont{and}
  \bibinfo{author}{\bibfnamefont{J.}~\bibnamefont{Zupan}},
  \bibinfo{journal}{JHEP} \textbf{\bibinfo{volume}{01}}, \bibinfo{pages}{051}
  (\bibinfo{year}{2014}), \eprint{1308.5663}.

\bibitem[{\citenamefont{Brod}(2015)}]{Brod:2014qwa}
\bibinfo{author}{\bibfnamefont{J.}~\bibnamefont{Brod}}, \bibinfo{journal}{Phys.
  Lett.} \textbf{\bibinfo{volume}{B743}}, \bibinfo{pages}{56}
  (\bibinfo{year}{2015}), \eprint{1412.3173}.

\bibitem[{\citenamefont{Patrignani et~al.}(2016)}]{Patrignani:2016xqp}
\bibinfo{author}{\bibfnamefont{C.}~\bibnamefont{Patrignani}}
  \bibnamefont{et~al.} (\bibinfo{collaboration}{Particle Data Group}),
  \bibinfo{journal}{Chin. Phys.} \textbf{\bibinfo{volume}{C40}},
  \bibinfo{pages}{100001} (\bibinfo{year}{2016}).

\bibitem[{\citenamefont{Bishara et~al.}(2018)\citenamefont{Bishara, Brod,
  Grinstein, and Zupan}}]{future:BBGZ}
\bibinfo{author}{\bibfnamefont{F.}~\bibnamefont{Bishara}},
  \bibinfo{author}{\bibfnamefont{J.}~\bibnamefont{Brod}},
  \bibinfo{author}{\bibfnamefont{B.}~\bibnamefont{Grinstein}},
  \bibnamefont{and} \bibinfo{author}{\bibfnamefont{J.}~\bibnamefont{Zupan}},
  \bibinfo{journal}{to appear}  (\bibinfo{year}{2018}).

\end{thebibliography}

\end{document}